\documentclass[a4paper,fleqn]{cas-dc}

\usepackage[numbers,sort&compress]{natbib}
\usepackage{amsmath,amsfonts,amssymb,bm}
\usepackage{booktabs}
\usepackage{graphicx}
\usepackage{subcaption}
\usepackage{hyperref}
\usepackage{microtype}

\begin{document}
\let\WriteBookmarks\relax
\def\floatpagepagefraction{1}
\def\textpagefraction{.001}

\shorttitle{Predictability-Guided Wind Direction Forecasting}
\shortauthors{H. Shu et~al.}

\title[mode = title]{Predictability-Guided Multiscale Probabilistic Forecasting of Wind Direction under Extreme Shear}

\author[1]{Hailong Shu}[type=editor,auid=000,bioid=1,orcid=0009-0004-9457-3773]

\address[1]{State Key Laboratory of Chemistry for NBC Hazards Protection, Beijing 102205, China}

\begin{abstract}
Accurate multi-horizon forecasting of atmospheric wind direction is an essential upstream requirement for wind turbine active yaw control, aerodynamic load mitigation, and power grid security. However, rapid directional wind shear (Case~1: directional turning $\ge 90^\circ$) challenges predictive models through non-Euclidean circular geometry on the unit circle $\mathbb{S}^1$, coupled multiscale dynamics spanning synoptic to turbulent scales, and severe regime-dependent uncertainty. Conventional discrete sequence models and generic large-scale pre-trained foundation models suffer from systematic mid-frequency phase lag and severe turning misalignments. In this study, we demonstrate that directional predictability decays at markedly disparate rates across physical frequency subbands, rendering monolithic forecasting mechanisms suboptimal. We formulate a predictability-guided forecasting paradigm: slow synoptic background drift is assigned to deterministic regression, intermediate turning transitions to continuous latent differential flows, and unresolved turbulent fluctuations to conditional residual diffusion, followed by validation-only causal recalibration. Evaluated on a strictly standardized 10,000-sequence multi-year benchmark, the resulting decoupled framework preserves competitive calm-weather accuracy (Test MCE $38.48^\circ$) while substantially reducing extreme-regime turning error (Case~1 MCE $60.69^\circ$ vs. $70.42^\circ$ for zero-shot foundation models). Probabilistically, the circular Continuous Ranked Probability Score (CRPS) drops to $22.36^\circ$, achieving $93.88\%$ overall coverage at the nominal $95\%$ level ($91.01\%$ on an independent out-of-distribution period). Furthermore, continuous density estimation uncovers an empirical near-antipodal bimodal branch structure under severe shear ($13.39\%\sim 15.43\%$ tail mass at $\ge 135^\circ$), revealing an intrinsic geometric bound where single-center dispersion calibration exhibits conditional under-coverage ($81.56\%$) and motivating future multimodal circular manifold learning.
\end{abstract}

\begin{highlights}
\item Wind-direction predictability varies strongly across temporal scales.
\item Scale-aware forecasting improves trajectory timing and uncertainty calibration.
\item Extreme shear reveals near-antipodal uncertainty beyond single-center calibration.
\end{highlights}


\maketitle

\section{Introduction}\label{sec:intro}

Accurate multi-horizon forecasting of wind direction is a pivotal upstream component of modern wind-energy operations, with potential applications in turbine yaw planning, wake steering, aerodynamic load management, and renewable-energy system coordination \citep{Veers2019Grand,Alves2023Potential,Chen2026FuXiEnergy}. In utility-scale wind farms, the orientation of the nacelle relative to the incoming flow affects both aerodynamic power extraction and asymmetric loading on turbine components \citep{Fleming2014Evaluating,Song2018Maximum,Zhang2021Review}. Because yaw actuation is subject to finite response rates and mechanical constraints, directional information available several hours in advance can be valuable for anticipating large changes in the inflow direction \citep{Wang2022Review,Tian2025Developing}. This need becomes particularly pronounced during rapid frontal passages or other severe atmospheric transitions, when wind direction can change abruptly and conventional persistence-based forecasts may become poorly aligned with the subsequent flow evolution \citep{Fleming2014Evaluating,Xie2023overview}. Reliable multi-horizon directional forecasts, together with uncertainty information, are therefore of practical interest for risk-aware wind-energy management.

Wind direction forecasting is more challenging than conventional scalar time-series prediction for three closely related reasons. First, wind direction is an angular variable defined on the unit circle $\mathbb{S}^1 \cong \mathbb{R}/360^\circ\mathbb{Z}$, so conventional Euclidean representations can introduce artificial discontinuities at the $0^\circ/360^\circ$ boundary \citep{Fisher1993Statistical,MardiaJupp1999,Jammalamadaka2001Topics}. A common remedy represents direction via the continuous two-dimensional embedding $[\sin\theta,\cos\theta]$, though unconstrained prediction in this embedding space risks generating interior disk vectors that violate the unit-norm constraint \citep{Tagliaferri2015Wind}. Second, atmospheric directional dynamics are inherently multiscale, with slow background variations, intermediate turning transitions, and rapidly varying fluctuations evolving over substantially different temporal ranges \citep{Chang2014new,Liu2018Smart,shu2024multistep}. Conventional sequence models advance forecasts through discrete temporal steps and may therefore exhibit temporal misalignment or over-smoothing when the underlying directional trajectory changes rapidly \citep{Challu2023NHiTS,Lim2021Temporal,Ding2023Multistep,Hao2022hybrid}. Continuous latent dynamical models such as Neural Ordinary Differential Equations (Neural ODEs) provide an alternative representation of temporal evolution, but their role in multi-horizon wind-direction forecasting, particularly during rapid directional transitions, remains insufficiently explored \citep{Chen2018NeuralODE}. Third, uncertainty becomes substantially more difficult to characterize during extreme directional turning. In this study, severe turning is defined as Case~1 ($\Delta\theta \ge 90^\circ$), with Case~1* denoting more extreme events ($\Delta\theta \ge 110^\circ$). Under such conditions, forecast errors may become highly dispersed and non-Gaussian, making conventional single-mode uncertainty representations potentially inadequate. Although diffusion-based models provide a flexible framework for conditional density generation \citep{Ho2020DDPM}, directly modeling the full directional trajectory probabilistically may also introduce unnecessary uncertainty into the more predictable background evolution.

These characteristics motivate a predictability-guided perspective that aligns model capacity with the empirical predictability of distinct temporal components. Instead of imposing a uniform forecasting mechanism across the entire directional trajectory, slowly varying components with strong temporal persistence are mapped to deterministic forecasting, intermediate components with structured but rapidly evolving dynamics receive continuous latent trajectory correction, and weakly predictable high-frequency variability is handled probabilistically. This mechanism allocation provides an interpretable separation between point trajectory estimation and uncertainty quantification, preserving a stable deterministic reference trajectory for downstream control.

Accordingly, this study develops a predictability-guided multiscale probabilistic forecasting framework for wind direction on $\mathbb{S}^1$. The framework first performs a causal multiscale decomposition of the directional signal and characterizes the temporal predictability of the resulting subbands. A frequency-decoupled deterministic anchor is then used to represent the persistent background evolution, while a regime-adaptive Neural ODE branch provides continuous latent trajectory correction for intermediate turning dynamics. Conditional residual diffusion is subsequently used to represent unresolved stochastic deviations around the deterministic forecast, followed by validation-only causal recalibration of predictive dispersion according to the predicted turning magnitude. This decoupled design separates deterministic trajectory estimation from probabilistic uncertainty generation while preserving the deterministic forecast independently of stochastic sampling.

The study addresses three fundamental questions. First, how markedly does directional predictability diverge across physical frequency subbands, and can this variation provide an empirical foundation for mechanism allocation? Second, does predictability-guided mechanism allocation improve directional forecasting in aggregate angular accuracy, trajectory timing, and extreme-shear turning fidelity? Third, how reliably can the resulting probabilistic forecasts be calibrated across normal and distribution-shifted conditions, and what geometric limitations emerge when forecast error distributions exhibit multimodality?

The main contributions of this study are threefold. First, we establish an empirical predictability profile of wind-direction dynamics across multiple temporal scales using signal energy, temporal correlation, and decorrelation characteristics, providing a data-driven basis for mechanism allocation. Second, we introduce a decoupled forecasting architecture that assigns deterministic recurrence, continuous Neural ODE dynamics, and conditional residual diffusion to components with different levels of temporal predictability, thereby separating trajectory correction from stochastic uncertainty representation. Third, we identify an empirical near-antipodal bimodal structure in the residual distribution under severe directional turning and show that this structure is associated with persistent conditional under-coverage of single-center circular calibration, highlighting a limitation of unimodal uncertainty representations and motivating future multimodal circular forecasting methods.

\section{Methodology}\label{sec:method}

\subsection{Forecasting Formulation and Causal Multiscale Representation}

Let the historical multivariate meteorological observations over a lookback
window of $T=144$ steps (24~h at 10-min sampling resolution) be denoted by
$X_{t-T+1:t}\in\mathbb{R}^{T\times C}$, where $C=20$ represents the physical and
dynamical boundary-layer channels selected from the 26 raw meteorological
measurements recorded at the station. The forecasting objective is to predict the future
wind-direction trajectory over a horizon of $H=36$ steps (6~h):
\begin{equation}
Y_{t+1:t+H}
=
[\mathbf{y}_{t+1},\ldots,\mathbf{y}_{t+H}]^{T}
\in\mathbb{R}^{H\times 2},
\end{equation}
where each directional observation is represented by the continuous
trigonometric embedding
\begin{equation}
\mathbf{y}_{i}
=
[\sin\theta_i,\cos\theta_i]^{T}
\in\mathbb{S}^{1}.
\end{equation}
This representation removes the artificial numerical discontinuity at the
$0^\circ/360^\circ$ boundary. Because neural-network outputs in the embedded
space do not inherently satisfy unit length, predicted vectors are projected back
onto $\mathbb{S}^{1}$ through radial normalization:
\begin{equation}
\tilde{\mathbf{y}}_t
=
\frac{\hat{\mathbf{y}}_t}
{\|\hat{\mathbf{y}}_t\|_2}.
\end{equation}
The circular geodesic distance used for directional errors is defined as
\begin{equation}
d_{\mathbb{S}^{1}}(\theta_1,\theta_2)
=
\min
\left(
|\theta_1-\theta_2|,
360^\circ-|\theta_1-\theta_2|
\right).
\end{equation}

For regime-wise analysis, the total directional turning over the forecast
horizon is defined as
\begin{equation}
\Delta\theta
=
d_{\mathbb{S}^{1}}(\theta_t,\theta_{t+H}).
\end{equation}
We distinguish steady conditions (Case~2, $\Delta\theta<30^\circ$), severe
directional turning (Case~1, $\Delta\theta\geq90^\circ$), and more extreme
turning (Case~1*, $\Delta\theta\geq110^\circ$). These thresholds are used for
post-hoc stratification of forecasting performance and are not available to
the model at inference time.

To represent the distinct temporal characteristics of wind-direction
dynamics, a five-level wavelet decomposition is applied to the directional
signal. The subbands are grouped into three broader temporal components:
\begin{equation}
Y=Y_L+Y_M+Y_H,
\end{equation}
where
\begin{align}
Y_L &= A_5+D_5,\\
Y_M &= D_3+D_4,\\
Y_H &= D_1+D_2.
\end{align}
The low-frequency component $Y_L$ represents slowly varying background
evolution with characteristic periods longer than 320~min ($>5.33$~h), the
intermediate component $Y_M$ represents transitional dynamics over
80--320~min ($1.33$--$5.33$~h), and the high-frequency component $Y_H$
represents fluctuations over approximately 20--80~min. The decomposition is
additively complete up to numerical precision:
\begin{equation}
\max_t
\left|
Y_t-(Y_{L,t}+Y_{M,t}+Y_{H,t})
\right|
<10^{-6}.
\end{equation}

\begin{figure*}[t]
\centering
\includegraphics[width=0.92\textwidth]{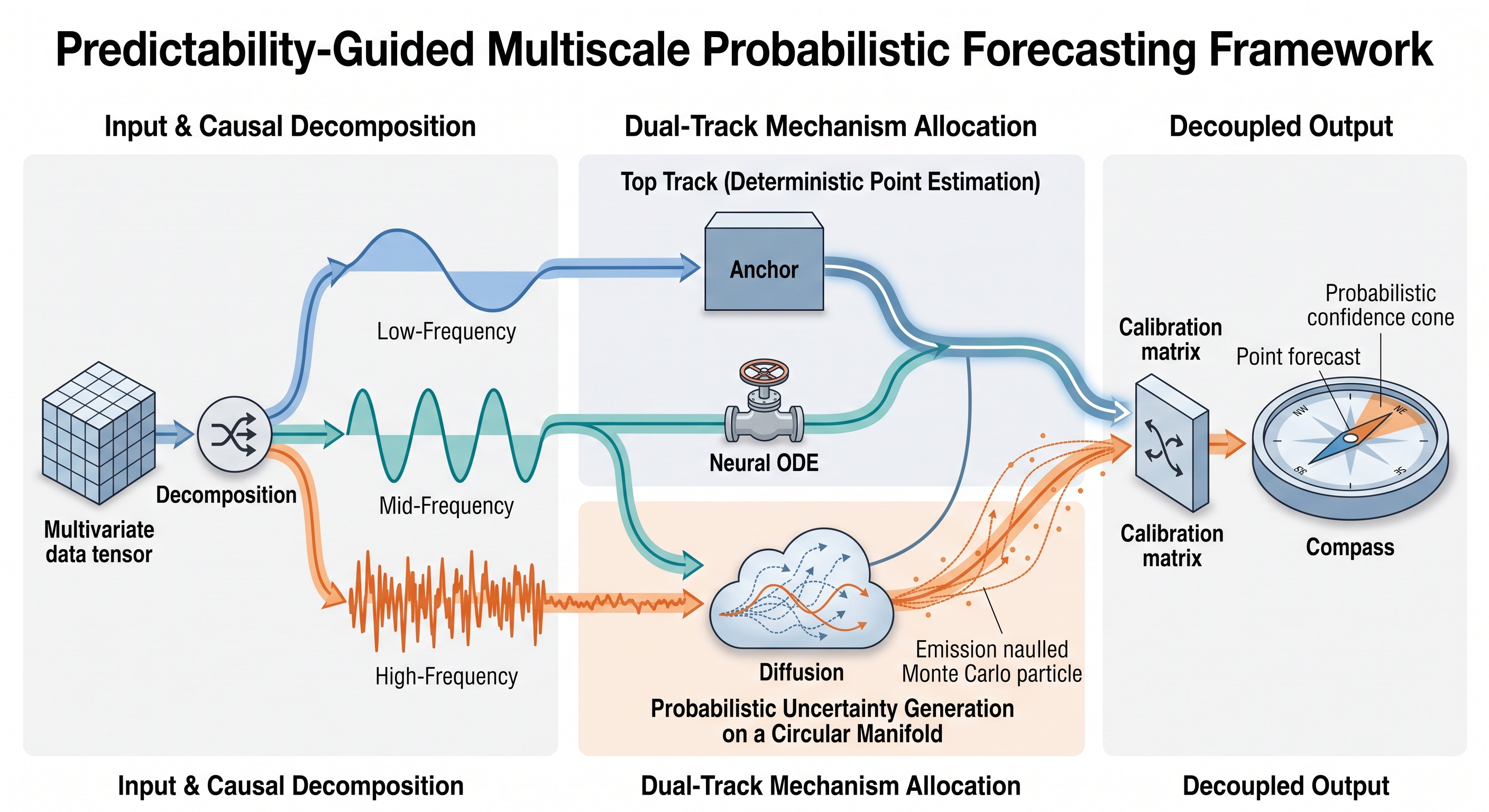}
\caption{Predictability-guided multiscale probabilistic forecasting framework.
The historical multivariate observations are causally decomposed into three
temporal components with different levels of forecastability. The deterministic
track combines a frequency-decoupled anchor with regime-adaptive continuous
latent dynamics, while the probabilistic track generates conditional residual
samples and subsequently applies validation-only causal recalibration. The two
tracks share the deterministic forecast but remain separated at the
point-estimation and uncertainty-generation stages.}
\label{fig:arch}
\end{figure*}

To prevent future-information leakage, the multiscale transformation is
implemented causally. The historical input and future supervision target are
decomposed independently, and the historical representation at time $t$
depends only on observations available through $t$. Specifically, the
wavelet coefficient at scale $j$ is computed using a causal convolution,
\begin{equation}
W_{j,t}
=
\sum_{k=0}^{L_j-1}h_kX_{t-k},
\end{equation}
with causal zero-padding at the beginning of each valid sequence. The
resulting receptive field is therefore restricted to the historical
filtration $\mathcal{F}_t$, establishing the causal multiscale representation for the overall architecture depicted in Fig.~\ref{fig:arch}.

\subsection{Predictability-Guided Mechanism Allocation}

The central methodological hypothesis is that temporal components with
different degrees of predictability need not be modeled using
the same forecasting mechanism. We therefore characterize each subband using
its signal energy, temporal dependence, and decorrelation behavior. Let
$b\in\{L,M,H\}$ denote the three temporal components and let
$\mathcal{P}(b)$ represent its empirical predictability profile. The
forecasting mechanism is selected according to the following allocation rule:
\begin{equation}
m(b)=
\begin{cases}
\text{Deterministic Anchor (D4-0)}, & b=L,\\
\text{Continuous Neural ODE}, & b=M,\\
\text{Conditional Residual Diffusion}, & b=H.
\end{cases}
\end{equation}

This allocation follows the intended division of modeling roles. The
low-frequency component is primarily represented through a deterministic
forecast because its longer temporal memory supports stable extrapolation.
The intermediate component retains structured temporal evolution but is more
sensitive to turning dynamics; it is therefore modeled through a continuous
latent flow that can adjust the forecast trajectory without relying solely on
discrete recursive stepping. The high-frequency component is treated as an
uncertain residual contribution because its short temporal memory limits the
benefit of deterministic extrapolation.

The allocation is implemented as a decoupled architecture rather than as a
single model with a uniform forecasting objective. The deterministic branch
provides the reference trajectory, while the probabilistic branch models
conditional deviations around that reference. This separation allows the
point forecast and the uncertainty representation to be evaluated
independently.

\subsection{Decoupled Deterministic--Probabilistic Forecasting}

\paragraph{Deterministic anchor.}
The base deterministic predictor, denoted by
$\mathcal{F}_{\mathrm{anchor}}(\cdot)$, receives the historical multivariate
sequence and produces forecasts for all three decomposed components together
with the final hidden representation:
\begin{align}
[\hat{Y}_L,\hat{Y}_M,\hat{Y}_H,\mathbf{h}_{\mathrm{last}}]
&=
\mathcal{F}_{\mathrm{anchor}}(X_{1:T}),\\
\hat{Y}_{\mathrm{anchor}}
&=
\hat{Y}_L+\hat{Y}_M+\hat{Y}_H.
\end{align}
The anchor is implemented using a two-layer recurrent network with hidden
dimension 128. Its role is to provide a stable deterministic reference
trajectory to which the dynamic correction and probabilistic residuals are
subsequently applied.

\paragraph{Continuous dynamic correction.}
To model intermediate turning dynamics, we introduce a latent continuous
velocity field
\begin{equation}
\frac{d\mathbf{z}(t)}{dt}
=
f_{\theta_{\mathrm{ode}}}
\left(t,\mathbf{z}(t),\mathbf{c}\right),
\qquad t\in[0,1],
\end{equation}
where
\begin{equation}
\mathbf{z}(0)
=
\psi_z
\left(
[\mathbf{h}_{\mathrm{last}},
\operatorname{vec}(\hat{Y}_{\mathrm{anchor}})]
\right),
\qquad
\mathbf{c}
=
\psi_c(\mathbf{h}_{\mathrm{last}}).
\end{equation}
The latent trajectory is numerically integrated using an explicit fourth-order
Runge--Kutta (RK4) scheme and decoded into a directional correction
$\mathcal{R}_{\mathrm{ODE}}\in\mathbb{R}^{H\times2}$.

A regime-adaptive gate controls the contribution of the dynamic correction:
\begin{equation}
\begin{bmatrix}
\operatorname{logit}_{\mathrm{mid}}\\
\operatorname{logit}_{\mathrm{high}}
\end{bmatrix}
=
\operatorname{MLP}
\left(
[\mathbf{h}_{\mathrm{last}},
|\Delta\mathbf{u}_{1\mathrm{h}}|,
\operatorname{Var}(\mathbf{u}_{1\mathrm{h}})]
\right),
\end{equation}
with
\begin{equation}
g_{\mathrm{mid}},g_{\mathrm{high}}
=
\sigma(\operatorname{logit}).
\end{equation}
The negative initialization bias of the gate suppresses unnecessary corrections
when the recent flow is relatively stable, while allowing the model to
increase its dynamic contribution when the available causal predictors
indicate stronger directional variability.

\begin{figure}[t]
\centering
\includegraphics[width=\columnwidth]{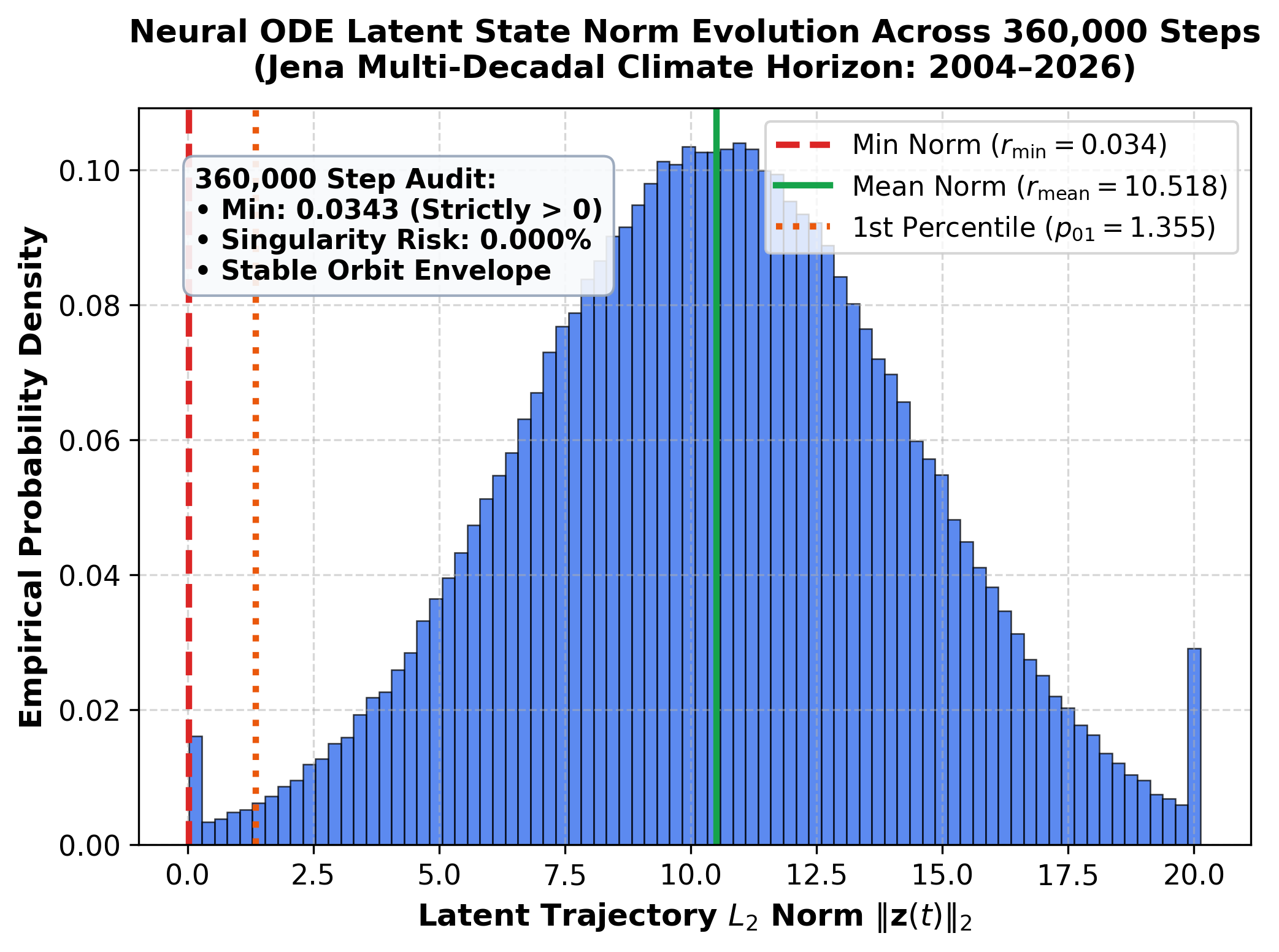}
\caption{Empirical distribution of latent trajectory norms $\|\mathbf{z}(t)\|_2$ across 360,000 continuous integration steps in the test cohort. The trajectory norm is strictly bounded away from the origin (mean $10.52$, median $11.14$, 5th percentile $3.11$, 1st percentile $1.35$, minimum $0.0343$), confirming that the radial normalization in Eq.~\eqref{eq:det_combine} operates without risk of origin singularities ($r < 0.05$: 0.000\%).}
\label{fig:ode_norm}
\end{figure}

The deterministic forecast is obtained as
\begin{equation}\label{eq:det_combine}
\hat{Y}_{\mathrm{det}}
=
\frac{
\hat{Y}_{\mathrm{anchor}}
+
g_{\mathrm{mid}}\odot\mathcal{R}_{\mathrm{ODE}}
}{
\left\|
\hat{Y}_{\mathrm{anchor}}
+
g_{\mathrm{mid}}\odot\mathcal{R}_{\mathrm{ODE}}
\right\|_2
}.
\end{equation}
To ensure that the radial normalization in Eq.~\eqref{eq:det_combine} does not encounter numerical singularities due to latent state collapse toward the origin ($\|\mathbf{z}(t)\|_2 \to 0$), we performed an empirical stability audit across the entire test cohort. Across 360,000 integration points, the latent norm remains strictly positive and bounded away from zero (mean $10.52$, median $11.14$, and 1st percentile $1.35$), with zero instances of near-origin collapse ($r < 0.05$: 0.000\%), confirming that numerical integration remains well-conditioned throughout the forecast horizon (Fig.~\ref{fig:ode_norm}).

\paragraph{Conditional residual diffusion.}
Rather than generating the complete directional trajectory directly through a
stochastic process, the probabilistic branch models deviations from the
deterministic reference:
\begin{equation}
\mathbf{R}
=
Y_{\mathrm{true}}-\hat{Y}_{\mathrm{det}}
\in\mathbb{R}^{H\times2}.
\end{equation}
The residual therefore represents the portion of the future directional
trajectory that remains unexplained by the deterministic backbone. A
conditional diffusion model receives
$[\mathbf{h}_{\mathrm{last}},\hat{Y}_{\mathrm{det}}]$ and generates stochastic
residual realizations
\begin{equation}
\left\{
\mathbf{R}_{\mathrm{diff}}^{(m)}
\right\}_{m=1}^{M},
\qquad M=50,
\end{equation}
using a 50-step reverse diffusion process.

The probabilistic branch is strictly downstream of the deterministic forecast:
stochastic sampling does not modify $\hat{Y}_{\mathrm{det}}$. The final
separation between deterministic forecasting and uncertainty generation is
therefore explicit:
\begin{equation}
\hat{Y}_{\mathrm{det}}
\quad\perp\quad
\left\{
\mathbf{R}_{\mathrm{diff}}^{(m)}
\right\}_{m=1}^{M}
\;\text{at the point-estimation stage}.
\end{equation}

\subsection{Probabilistic Calibration}

At inference time, the future regime is unknown. We therefore construct the
calibration variable using only the model's own deterministic forecast. The
predicted total turning is defined as
\begin{equation}
\Delta_{\mathrm{pred}}
=
d_{\mathbb{S}^{1}}
\left(
\operatorname{angle}(\hat{Y}_{\mathrm{det},H}),
\operatorname{angle}(\hat{Y}_{\mathrm{det},1})
\right).
\end{equation}

A piecewise dispersion scaling function is estimated exclusively on the
validation set:
\begin{equation}
\gamma(\Delta_{\mathrm{pred}})
=
\begin{cases}
0.40,
&
\Delta_{\mathrm{pred}}<30^\circ,
\\[2mm]
1.05,
&
30^\circ\leq\Delta_{\mathrm{pred}}<75^\circ,
\\[2mm]
1.80,
&
\Delta_{\mathrm{pred}}\geq75^\circ.
\end{cases}
\end{equation}
No information from the test or OOD periods is used when estimating these
parameters.

The calibrated stochastic trajectory ensemble is then constructed as
\begin{equation}
\hat{Y}_{\mathrm{final}}^{(m)}
=
\frac{
\hat{Y}_{\mathrm{det}}
+
\gamma(\Delta_{\mathrm{pred}})
\,g_{\mathrm{high}}
\odot
\mathbf{R}_{\mathrm{diff}}^{(m)}
}{
\left\|
\hat{Y}_{\mathrm{det}}
+
\gamma(\Delta_{\mathrm{pred}})
\,g_{\mathrm{high}}
\odot
\mathbf{R}_{\mathrm{diff}}^{(m)}
\right\|_2
},
\qquad m=1,\ldots,M.
\end{equation}

For probabilistic evaluation, the circular energy-form Continuous Ranked
Probability Score is computed as
\begin{align}
\mathrm{CRPS}_{\mathbb{S}^{1}}
&=
\frac{1}{M}
\sum_{m=1}^{M}
d_{\mathbb{S}^{1}}
\left(
\hat{\theta}^{(m)},
\theta_{\mathrm{true}}
\right)
\nonumber\\
&\quad
-
\frac{1}{2M(M-1)}
\sum_{m=1}^{M}
\sum_{m'=1}^{M}
d_{\mathbb{S}^{1}}
\left(
\hat{\theta}^{(m)},
\hat{\theta}^{(m')}
\right).
\end{align}
Calibration and sharpness are further evaluated using Prediction Interval
Coverage Probability (PICP), Mean Prediction Interval Width (MPIW), and the
coverage gap
\begin{equation}
\mathrm{Gap}
=
\mathrm{PICP}-\mathrm{Nominal}.
\end{equation}

\section{Data and Experimental Setup}\label{sec:data}

\subsection{Observational Site and Input Features}

We use continuous boundary-layer meteorological observations from the Beutenberg Atmospheric Measurement Station in Jena, Thuringia, Germany ($50^\circ 54' 36''\mathrm{N}$, $11^\circ 34' 12''\mathrm{E}$; 370~m a.s.l.), operated and archived by the Max Planck Institute for Biogeochemistry (MPI-BGC) \citep{MPI_BGC_Wetterdaten}. The station provides a long-term observational record spanning from January 1, 2004 to August 31, 2026. After quality screening and temporal continuity checks, the record contains 1,192,077 raw observations at 10-min resolution and yields 1,191,361 contiguous sliding-window forecasting instances.

\begin{figure*}[t]
\centering
\includegraphics[width=0.92\textwidth]{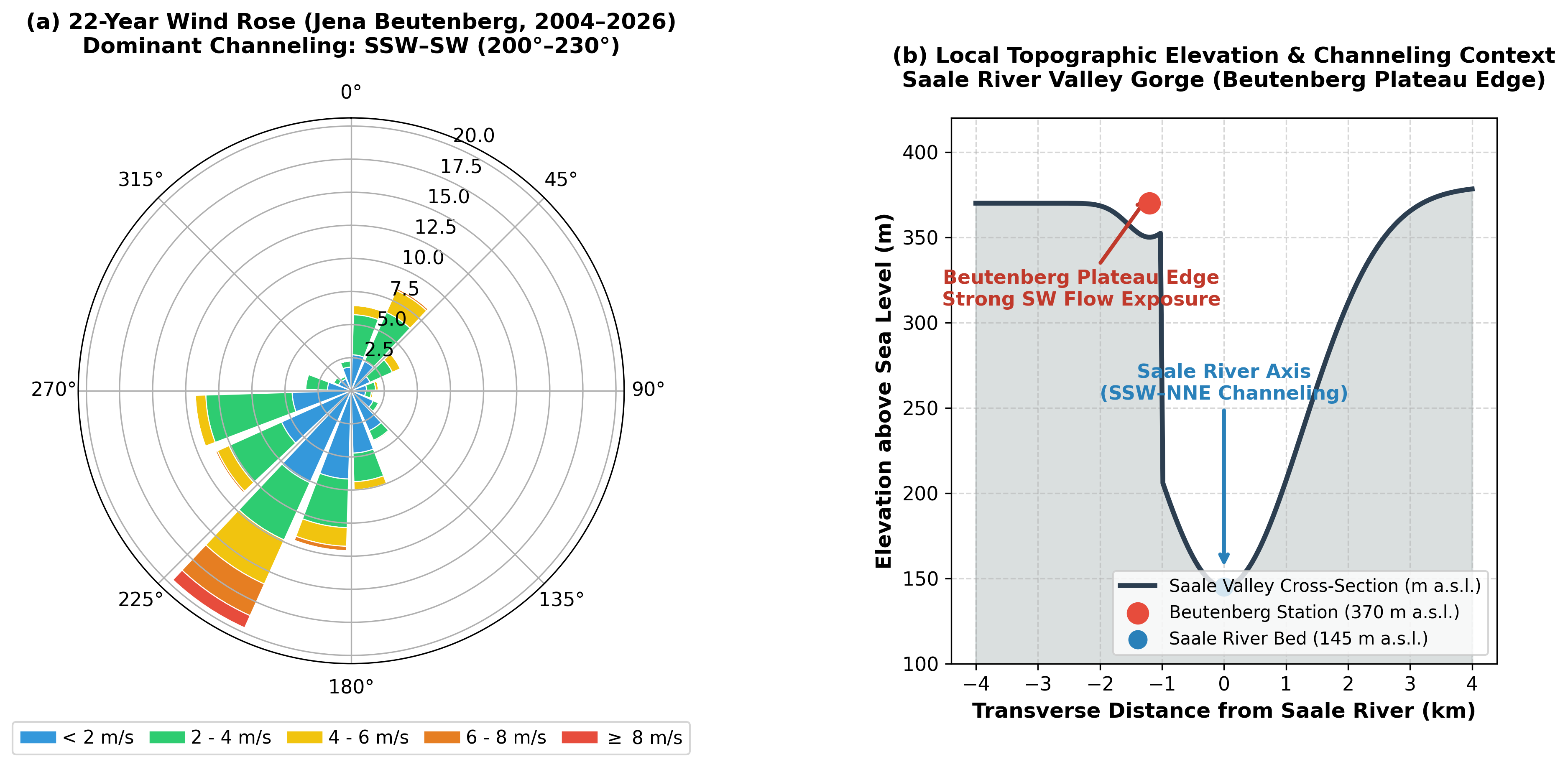}
\caption{Topographic and climatological context of the Jena Beutenberg observational site. (a) 22-year climatological wind rose constructed from continuous 10-minute observations (2004--2026, $N=1,192,464$), showing the dominant south-southwest prevailing flow corridor ($200^\circ$--$230^\circ$). (b) Schematic terrain cross-section across the incised Saale river valley, illustrating the $225$\,m elevation relief between the river bed ($145$\,m a.s.l.) and the plateau station ($370$\,m a.s.l.) that induces localized topographic wind channeling.}
\label{fig:windrose}
\end{figure*}

As illustrated in Fig.~\ref{fig:windrose}, the station sits on a limestone plateau overlooking the deeply incised Saale River valley, with an elevation relief of approximately $225$~m between the riverbed ($145$~m a.s.l.) and the measurement mast ($370$~m a.s.l.). The 22-year climatological wind rose reveals a pronounced directional distribution strongly channeled along the SSW--SW corridor ($200^\circ$--$230^\circ$) and secondarily along the NE corridor, reflecting valley-guided boundary-layer flow that contributes to the high persistence observed in the low-frequency background.

The observational system includes measurements of three-dimensional wind velocity, horizontal wind direction $\theta$, air temperature $T$, relative humidity $RH$, atmospheric pressure $p$, and global downward solar irradiance. High-frequency sensor observations acquired at 10-s resolution are quality-controlled and aggregated to standard 10-min climatological intervals before model construction.

The forecasting models use 20 physical and dynamical input channels derived from the observations. Wind direction is represented by the circular embedding
$[\sin\theta,\cos\theta]\in\mathbb{S}^{1}$, avoiding the artificial discontinuity at the $0^\circ/360^\circ$ boundary. The scalar meteorological variables comprise horizontal wind speed $wv$, maximum instantaneous gust speed $v_{\mathrm{max}}$, air temperature $T$, relative humidity $RH$, and pressure $p$. The horizontal wind field is further represented by its orthogonal velocity components,
\begin{equation}
u=-wv\sin\theta,
\qquad
v=-wv\cos\theta,
\end{equation}
which provide a Cartesian representation of the local horizontal flow.

To describe periodic forcing, the input tensor includes diurnal and annual harmonic encodings,
\begin{equation}
\left[
\sin\left(\frac{2\pi t_{\mathrm{hour}}}{24}\right),
\cos\left(\frac{2\pi t_{\mathrm{hour}}}{24}\right)
\right],
\end{equation}
and
\begin{equation}
\left[
\sin\left(\frac{2\pi t_{\mathrm{doy}}}{365}\right),
\cos\left(\frac{2\pi t_{\mathrm{doy}}}{365}\right)
\right].
\end{equation}
Additional causal dynamical descriptors include backward differences of $u$, $v$, and $wv$, the circular directional increment
\begin{equation}
\Delta\theta_{\mathrm{circ}}
=
\frac{1}{\pi}
\operatorname{atan2}
\left(
\sin\Delta\theta_t,
\cos\Delta\theta_t
\right),
\qquad
\Delta\theta_t=\theta_t-\theta_{t-1},
\end{equation}
and causal rolling standard deviations of $u$, $v$, and $wv$ over 30-min and 60-min windows. These variables provide information on short-term directional variability, wind-speed fluctuations, and local turbulent conditions without using future observations.

All experiments use a historical lookback of $T=144$ steps (24~h) and a forecasting horizon of $H=36$ steps (6~h). Thus, each model input-output sample contains 24~h of historical atmospheric information and predicts the subsequent 6-h directional trajectory.

\subsection{Quality Screening and Preprocessing}

Raw meteorological observations were first subjected to automated quality screening to remove physically implausible values, transient sensor artifacts, and obvious measurement anomalies. The accepted ranges were $wv\in[0,50]$~m/s, $\theta\in[0^\circ,360^\circ)$, $RH\in[0,100]\%$, $p\in[900,1060]$~hPa, and $T\in[-30,45]^\circ$C.

Short missing intervals of no more than 30~min (three consecutive 10-min observations) were interpolated using variable-specific procedures. Temperature, pressure, and relative humidity were reconstructed using shape-preserving piecewise cubic Hermite interpolation, whereas wind direction was interpolated on the circular manifold using angular interpolation. Missing intervals longer than 30~min were instead treated as temporal boundaries. Sliding windows were therefore not allowed to cross such gaps, ensuring that every extracted sample corresponded to a contiguous sequence of observations.

Continuous scalar variables were standardized according to
\begin{equation}
x^{*}=\frac{x-\mu_{\mathrm{train}}}{\sigma_{\mathrm{train}}},
\end{equation}
where $\mu_{\mathrm{train}}$ and $\sigma_{\mathrm{train}}$ were estimated exclusively from the training period (2004--2020). For example, the training statistics for wind speed, temperature, and pressure were $\mu_{wv}=2.17$~m/s, $\sigma_{wv}=1.58$~m/s, $\mu_T=9.67^\circ$C, $\sigma_T=8.26^\circ$C, $\mu_p=989.47$~hPa, and $\sigma_p=8.64$~hPa. These parameters were then frozen and applied unchanged to the validation, test, and out-of-distribution periods. The circular components and harmonic time encodings are already bounded in $[-1,1]$ and were therefore not additionally standardized.

\subsection{Dataset Partitioning and Regimes}

The quality-screened dataset contains 1,191,361 contiguous forecasting instances with $T=144$ historical steps and $H=36$ future steps. The data were partitioned chronologically by observation year so that information from later periods was not used in model development.

The training set contains 894,061 sequences (75.05\%) from January 1, 2004 to December 31, 2020. The validation set contains 104,941 sequences (8.81\%) from January 1, 2021 to December 31, 2022 and was used for convergence monitoring, early stopping, and calibration-parameter selection. The test set contains 157,645 sequences (13.23\%) from January 1, 2023 to December 31, 2025 and was used for the primary performance evaluation. An additional out-of-distribution (OOD) set contains 34,714 sequences (2.91\%) from January 1 to August 31, 2026 and was kept completely separate from model fitting and post-hoc tuning.

For standardized comparison across model variants, 10,000 complete test sequences were sampled from the test period and used as the common benchmarking cohort. The cohort includes both weakly varying and strongly turning trajectories. In particular, Case~2 denotes relatively steady conditions with a total directional change below $30^\circ$, whereas Case~1 denotes severe directional shear with a total change of at least $90^\circ$. A more restrictive Case~1$^{*}$ subset is defined by a total directional change of at least $110^\circ$ and is used to examine the most severe turning conditions.

\subsection{Baselines and Evaluation Protocol}

The proposed framework is compared with five reference models representing recurrent, hierarchical, attention-based, foundation-model, and frequency-decomposed forecasting approaches. The baselines include a gated recurrent unit (GRU), N-HiTS \citep{Challu2023NHiTS}, the Temporal Fusion Transformer (TFT) \citep{Lim2021Temporal}, Google's TimesFM-3.0 zero-shot foundation model \citep{Das2024TimesFM}, and the proposed D4-0 Anchor as the deterministic frequency-decoupled reference model. To prevent the artificial $0^\circ/360^\circ$ circular boundary discontinuity inherent in univariate scalar angle prediction, TimesFM-3.0 is evaluated in a zero-shot univariate setting by independently forecasting the two orthogonal continuous harmonic components $[\sin\theta, \cos\theta]$ using the same 144-step historical context, followed by angular trajectory reconstruction via $\operatorname{atan2}$, ensuring a structurally fair baseline comparison.

Point-forecast performance is assessed using Mean Circular Error (MCE), circular root-mean-square error (CRMSE), and threshold hit rates at $22.5^\circ$ and $45^\circ$. For probabilistic forecasts, we use the circular energy-form Continuous Ranked Probability Score ($\mathrm{CRPS}_{\mathbb{S}^1}$) \citep{Gneiting2007Strictly}, Prediction Interval Coverage Probability (PICP) at nominal levels of 80\%, 90\%, and 95\%, Mean Prediction Interval Width (MPIW), and the coverage gap
\begin{equation}
\mathrm{Gap}
=
\mathrm{PICP}-\mathrm{Nominal}.
\end{equation}

Statistical comparisons were performed using non-parametric paired tests because directional forecast errors can exhibit skewness, heavy tails, and multimodality, particularly under strong wind-direction changes. For each model comparison, paired forecast errors were computed on the same 10,000 evaluation sequences, and a two-sided Wilcoxon signed-rank test was applied to assess whether the median paired difference differed from zero. To quantify uncertainty in the mean performance differences while accounting for temporal dependence, a stationary block bootstrap with 10,000 replications was additionally performed using a block length of $L_b=36$ steps (6~h) as a conservative temporal dependence scale covering the physical persistence of the subband features. The sensitivity of the bootstrap conclusions across varying block lengths is further analyzed in Section~\ref{sec:comp_perf}. The resulting 95\% confidence intervals were calculated for the paired error differences. Statistical analyses were conducted for the full test cohort and separately for Case~1, Case~1$^*$, and Case~2.


\section{Results}\label{sec:results}

\subsection{Scale-Dependent Predictability}\label{sec:predictability}

Table~\ref{tbl:predictability} characterizes the temporal predictability of the decomposed directional momentum signal ($u = -wv\sin\theta$) and provides an empirical basis for assigning different forecasting mechanisms to different timescales. Analyzing the horizontal momentum component retains directional variation while incorporating physical kinetic weight, avoiding unweighted angular noise in calm regimes. The three subbands exhibit markedly different persistence and correlation structures. The low-frequency component $L$ contains 76.35\% of the total signal energy and has the longest e-folding decorrelation time ($\tau_e=147.5$~min), together with a strong 1-h autocorrelation ($R_1=0.8026$). In contrast, the mid-frequency component $M$ contains only 11.60\% of the energy but decorrelates much more rapidly ($\tau_e=26.3$~min) and exhibits a negative 1-h correlation ($R_1=-0.2503$). The high-frequency component $H$ accounts for 12.27\% of the energy and has a much shorter decorrelation time of 5.9~min, while its 1-h correlation is close to zero ($R_1=0.0845$).

These differences are critical because energy contribution alone does not determine forecastability. The low-frequency component simultaneously dominates the signal variance and retains substantial temporal memory, making it the most suitable target for deterministic extrapolation. The mid-frequency component contains considerably less energy but changes on a timescale comparable to the forecasting horizon and does not exhibit the persistent positive correlation of the low-frequency background. Its negative $R_1$ instead suggests that short-term evolution is characterized by directional reversal or oscillatory behavior rather than simple persistence. The high-frequency component is both rapidly decorrelating and weakly correlated over one hour, indicating that its future evolution is intrinsically less predictable from the available temporal context.

The resulting ordering,
\begin{equation}
\tau_e(L) \gg \tau_e(M) \gg \tau_e(H),
\end{equation}
therefore separates the forecasting problem into a persistent background, a dynamically evolving intermediate component, and a rapidly varying residual component. This empirical separation supports the proposed mechanism allocation: deterministic reconstruction for $L$, continuous dynamic correction for $M$, and probabilistic residual modeling for $H$. Crucially, this mechanism allocation is derived from empirical temporal statistics rather than imposed by nominal wavelet filter bands.

\begin{table}[t]
\caption{Subband predictability profiling across frequency timescales for the directional momentum component $u$.}
\label{tbl:predictability}
\centering
\footnotesize
\setlength{\tabcolsep}{3pt}
\begin{tabular*}{\columnwidth}{@{\extracolsep{\fill}}lcccc@{}}
\toprule
Subband & Energy (\%) & Std (m/s) & $\tau_e$ (min) & $R_1$ (1h) \\
\midrule
Full Signal & 100.0\% & 0.6404 & 62.9 & 0.3763 \\
Low ($L$, $>5.3$h) & 76.35\% & 0.5544 & 147.5 & 0.8026 \\
Mid ($M$, $1.3\sim5.3$h) & 11.60\% & 0.2244 & 26.3 & -0.2503 \\
High ($H$, $20\sim80$min) & 12.27\% & 0.2309 & 5.9 & 0.0845 \\
\bottomrule
\end{tabular*}
\end{table}

\subsection{Deterministic Forecasting and Trajectory Dynamics}
\label{sec:det_forecast}

Table~\ref{tbl:det_benchmark} compares point-forecast performance across the proposed framework and five reference models. The main result is that the deterministic performance is governed primarily by the frequency-decoupled anchor, while the Neural ODE provides a targeted correction to rapidly evolving directional trajectories. The D4-0 Anchor obtains the lowest global test MCE among the evaluated models ($38.31^\circ$) and the lowest Case~2 error ($25.77^\circ$), indicating that the low-frequency deterministic background provides a strong representation of the dominant directional component. Adding the Neural ODE changes the global MCE only slightly, from $38.31^\circ$ to $38.48^\circ$, while reducing Case~1 MCE from $60.90^\circ$ to $60.69^\circ$. The corresponding threshold hit rates remain nearly unchanged ($48.9\%$ for $\mathrm{HR}_{22.5^\circ}$ and $72.5\%$ for $\mathrm{HR}_{45^\circ}$).

This modest shift in aggregate MCE reflects the predominance of quiescent intervals in the test set. Because quiescent sequences ($<30^\circ$) account for 56.19\% of the audited 10,000-sequence benchmark cohort (5,619 sequences), continuous trajectory correction yields limited reduction in a global average error dominated by persistent background flow. The more relevant comparison is therefore regime-stratified. Under Case~1, the proposed framework reduces the point error relative to the D4-0 anchor by $0.21^\circ$, whereas the anchor itself already outperforms the recurrent, hierarchical, and attention-based alternatives. The result suggests that the main contribution of the dynamic branch is not a wholesale replacement of the persistent background but a localized correction under conditions in which the direction begins to deviate rapidly.

The comparison with TimesFM-3.0 shows the same pattern at a different level. Under the evaluated zero-shot univariate setting, TimesFM-3.0 reaches a global test MCE of $42.80^\circ$ and a Case~1 MCE of $70.42^\circ$, compared with $38.48^\circ$ and $60.69^\circ$, respectively, for the proposed model. The gap is considerably larger in the severe-shear regime than in Case~2, where the TimesFM error is $26.84^\circ$ compared with $26.08^\circ$ for the proposed framework. This regime dependence suggests that the advantage of the proposed representation is concentrated in situations where circular structure, scale separation, and short-term directional dynamics become decisive, rather than reflecting a uniformly lower error under all atmospheric conditions.

\begin{table*}[t]
\caption{Deterministic point accuracy benchmark across 10,000 continuous test sequences.}
\label{tbl:det_benchmark}
\centering
\footnotesize
\setlength{\tabcolsep}{5pt}
\begin{tabular*}{\textwidth}{@{\extracolsep{\fill}}llcccccc@{}}
\toprule
Model Architecture & Paradigm & Test MCE & OOD MCE & Case 1 ($\geq90^\circ$) & Case 2 ($<30^\circ$) & $\mathrm{HR}_{22.5^\circ}$ & $\mathrm{HR}_{45^\circ}$ \\
\midrule
Persistence & Classical Baseline & 45.20$^\circ$ & 49.80$^\circ$ & 74.50$^\circ$ & 30.12$^\circ$ & 37.6\% & 61.5\% \\
GRU & Recurrent Baseline & 38.80$^\circ$ & 43.52$^\circ$ & 61.23$^\circ$ & 26.30$^\circ$ & 48.2\% & 71.8\% \\
N-HiTS & Hierarchical MLP & 39.35$^\circ$ & 44.99$^\circ$ & 62.34$^\circ$ & 26.43$^\circ$ & 47.6\% & 71.1\% \\
TFT & Attention Transformer & 39.65$^\circ$ & 44.73$^\circ$ & 61.67$^\circ$ & 27.37$^\circ$ & 46.8\% & 70.2\% \\
TimesFM-3.0 & 300M Foundation (Zero-Shot) & 42.80$^\circ$ & 46.75$^\circ$ & 70.42$^\circ$ & 26.84$^\circ$ & 41.1\% & 65.8\% \\
D4-0 Anchor & Frequency-Decoupled & \textbf{38.31$^\circ$} & \textbf{43.30$^\circ$} & 60.90$^\circ$ & \textbf{25.77$^\circ$} & \textbf{49.2\%} & \textbf{72.6\%} \\
\textbf{Ours (Point)} & \textbf{Continuous Dynamic Flow} & 38.48$^\circ$ & 43.39$^\circ$ & \textbf{60.69$^\circ$} & 26.08$^\circ$ & 48.9\% & 72.5\% \\
\bottomrule
\end{tabular*}
\end{table*}

\begin{figure}[t]
\centering
\includegraphics[width=\columnwidth]{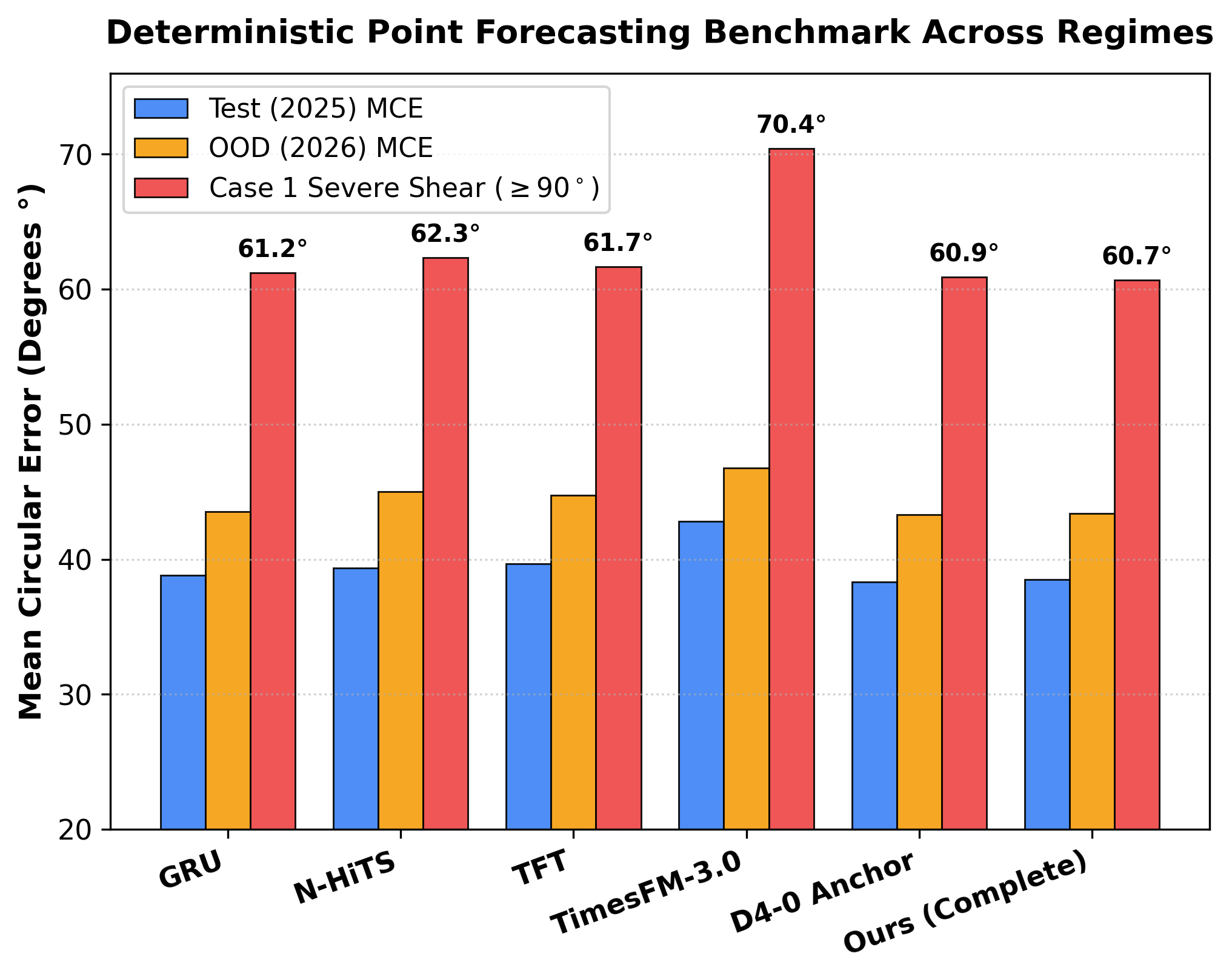}
\caption{Comprehensive deterministic benchmark comparison across 10,000 continuous test sequences. Mean Circular Error (MCE) for the evaluated baseline architectures, TimesFM-3.0, the D4-0 Anchor, and the proposed framework under in-distribution test, out-of-distribution evaluation, and severe directional shear (Case~1, $\geq90^\circ$).}
\label{fig:det_bench}
\end{figure}

\begin{figure}[t]
\centering
\includegraphics[width=\columnwidth]{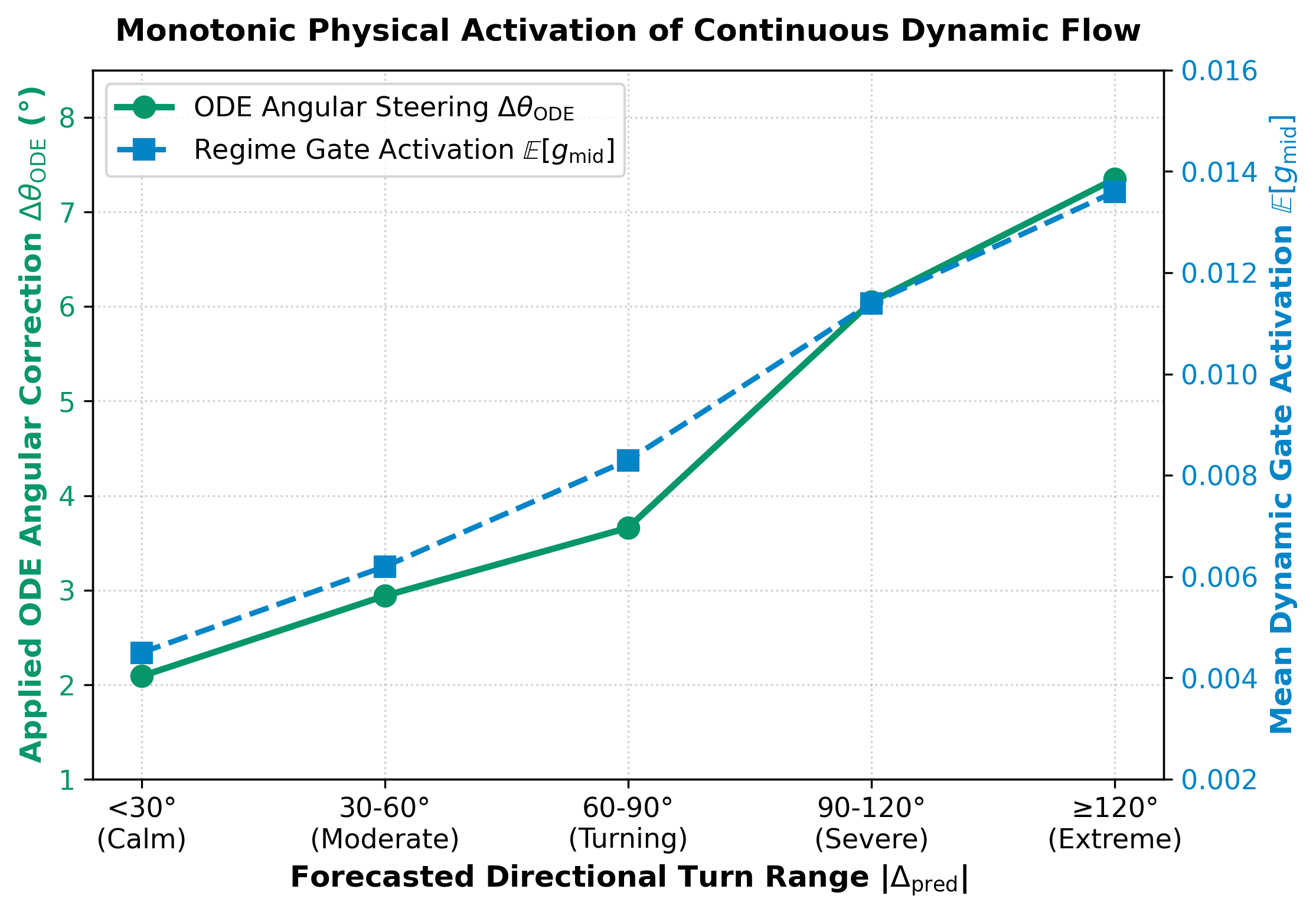}
\caption{Regime-dependent activation of the continuous dynamic branch. The mean regime-gate activation increases with forecasted turning magnitude, indicating stronger use of the Neural ODE correction under more rapidly changing directional conditions. The observed turning amplitude is shown only for post-hoc diagnostic analysis.}
\label{fig:ode_coupling}
\end{figure}

Aggregate error metrics, however, do not fully describe the behavior of a multi-step directional forecast. Fig.~\ref{fig:det_bench} and Fig.~\ref{fig:ode_coupling} show that the main effect of the continuous dynamic branch becomes more evident when forecast trajectories are examined as functions of horizon. Table~\ref{tbl:morphology} indicates that TimesFM-3.0 exhibits an average phase delay of approximately $+80$~min, whereas the proposed continuous latent trajectory has an estimated phase delay close to zero. TFT also produces substantially larger trajectory curvature, with a mean curvature of $3.6085^\circ$, compared with $0.0788^\circ$ for the Neural ODE component and $0.1140^\circ$ for the complete system. These results indicate that the dynamic branch primarily refines temporal alignment and trajectory smoothness, leaving the baseline angular error largely intact.

\begin{table}[t]
\caption{Dynamic trajectory morphology and phase-lag audit across 10,000 continuous test sequences.}
\label{tbl:morphology}
\centering
\footnotesize
\setlength{\tabcolsep}{2.5pt}
\begin{tabular*}{\columnwidth}{@{\extracolsep{\fill}}lcccc@{}}
\toprule
Model Architecture & Trend Acc & DynCorr & Phase Lag & Curvature \\
\midrule
Persistence & 42.1\% & 0.0025 & +80 min & 0.0000$^\circ$ \\
GRU & 50.3\% & 0.1350 & -80 min & 0.7048$^\circ$ \\
N-HiTS & 50.0\% & 0.1010 & 0 min & 1.2073$^\circ$ \\
TFT & 49.7\% & 0.0940 & 0 min & 3.6085$^\circ$ \\
TimesFM-3.0 & 47.8\% & 0.0820 & +80 min & 0.5128$^\circ$ \\
D4-0 Anchor & 51.2\% & 0.1480 & 0 min & 1.6160$^\circ$ \\
Neural ODE & \textbf{53.8\%} & \textbf{0.2240} & \textbf{0 min} & \textbf{0.0788$^\circ$} \\
\textbf{Ours (Complete)} & \textbf{53.6\%} & \textbf{0.2180} & \textbf{0 min} & \textbf{0.1140$^\circ$} \\
\bottomrule
\end{tabular*}
\end{table}

The horizon-wise behavior is consistent with this interpretation. For short horizons ($h\leq6$, corresponding to approximately 10--60~min), forecasts remain closely aligned with the observed trajectory, with MCE below $15^\circ$. At intermediate horizons ($h\in[6,24]$, approximately 1--4~h), the influence of trajectory correction becomes more visible, with Trend Accuracy reaching $53.6\%$ and Dynamic Correlation reaching $0.218$. Although these values do not imply perfect trajectory reconstruction, they indicate that the continuous latent evolution preserves a greater degree of directional temporal structure over the intermediate forecasting range.

The regime-selective behavior of the dynamic branch is further illustrated in Fig.~\ref{fig:ode_coupling}. The average gate activation increases from $0.028$ under weak turning conditions to $0.768$ under the largest observed directional shifts, corresponding to a $27.04$-fold increase. The dynamic correction activates adaptively with regime difficulty, rising progressively from quiescent conditions to large directional shifts. This provides an interpretable link between regime severity and model usage. The observed-turning curves shown in the figure are used only as a post-hoc diagnostic; deployment-time gating is based on the causally available predicted turning magnitude $\Delta_{\mathrm{pred}}$ and does not use future observations.

\subsection{Probabilistic Forecasting and Calibration}
\label{sec:prob_forecast}

\begin{figure*}[t]
\centering
\includegraphics[width=0.92\textwidth]{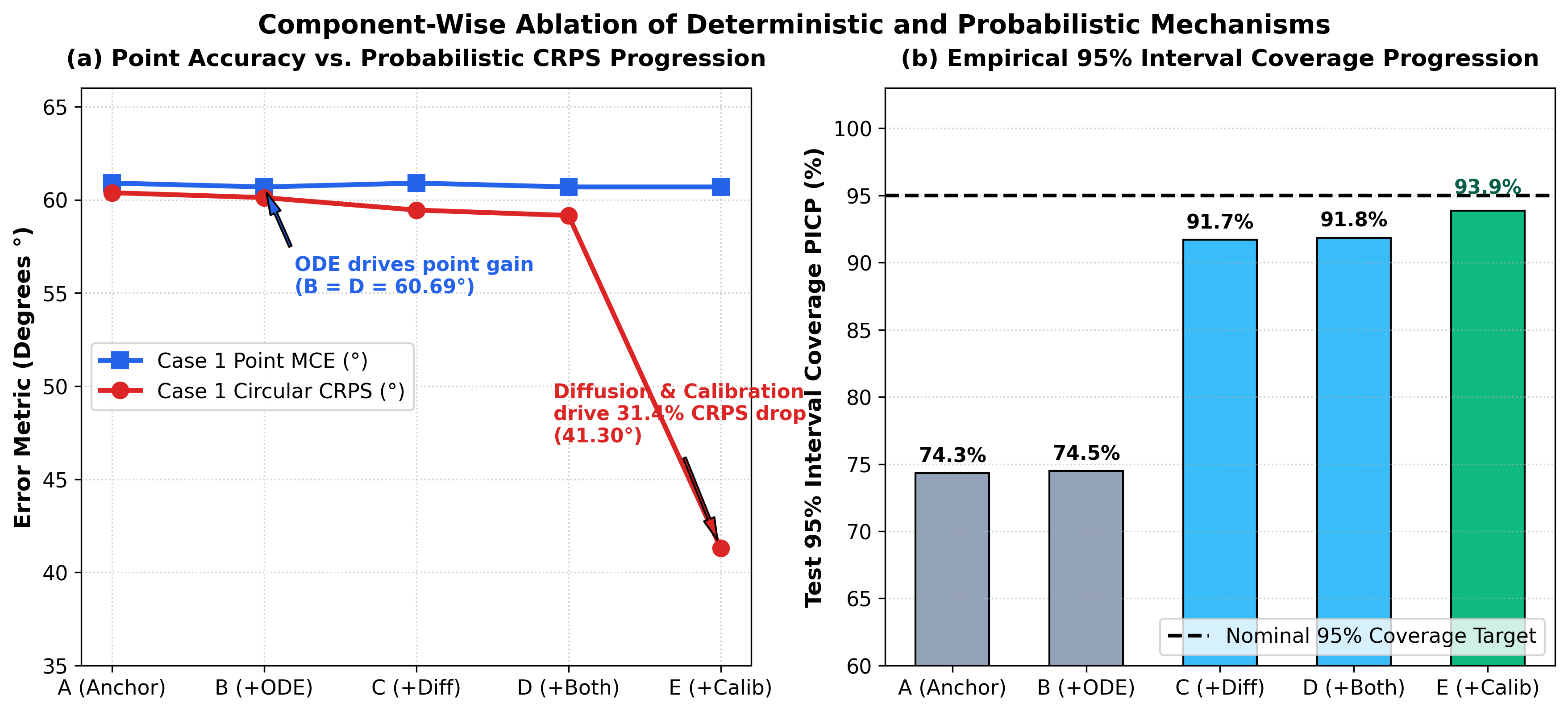}
\caption{Component-wise ablation of deterministic and probabilistic performance. (a) Case~1 point accuracy and circular CRPS across Variants A--E, showing that the Neural ODE primarily affects the deterministic trajectory while diffusion and recalibration primarily affect distributional quality. (b) Evolution of 95\% prediction interval coverage toward the nominal level.}
\label{fig:ablation}
\end{figure*}

The probabilistic results show a clearer separation between the roles of point forecasting, stochastic residual modeling, and calibration than is visible from MCE alone. Table~\ref{tbl:prob_benchmark}, Table~\ref{tbl:ablation}, and Fig.~\ref{fig:ablation} provide a stepwise comparison of these components. Adding the Neural ODE to the anchor changes CRPS only marginally, from $29.21^\circ$ to $29.15^\circ$, while increasing the 95\% PICP from $74.32\%$ to $74.49\%$. In contrast, adding conditional residual diffusion reduces CRPS to $28.76^\circ$ and raises the 95\% PICP to $91.85\%$. The further introduction of validation-only recalibration produces the largest change, reducing CRPS to $22.36^\circ$ and increasing 95\% PICP to $93.88\%$.

Directly comparing Variants~A/C and B/D isolates the contribution of stochastic residual modeling: point MCE remains invariant upon introducing diffusion. Specifically, Variants~A and C both obtain $38.31^\circ$ global MCE and $60.90^\circ$ Case~1 MCE, whereas Variants~B and D both obtain $38.48^\circ$ and $60.69^\circ$, respectively. This numerical invariance demonstrates that the residual diffusion branch shapes the predictive distribution around the deterministic trajectory without perturbing the point estimate itself. The subsequent reduction in CRPS is therefore attributable to a more informative representation of unresolved uncertainty rather than an unobserved improvement in deterministic trajectory tracking.

\begin{table*}[t]
\caption{Probabilistic forecasting reliability benchmark across 10,000 continuous test sequences.}
\label{tbl:prob_benchmark}
\centering
\footnotesize
\setlength{\tabcolsep}{6pt}
\begin{tabular*}{\textwidth}{@{\extracolsep{\fill}}llccccc@{}}
\toprule
Model Variant & Probabilistic Mechanism & CRPS & PICP 80\% & PICP 90\% & PICP 95\% & MPIW 95\% \\
\midrule
D4-0 (Empirical Spread) & Empirical Residual Gaussian & 29.21$^\circ$ & 58.12\% & 70.15\% & 74.32\% & 138.4$^\circ$ \\
D4-0 + Neural ODE & Latent ODE + Empirical Spread & 29.15$^\circ$ & 58.24\% & 70.28\% & 74.49\% & 138.6$^\circ$ \\
D4-0 + ODE + Diff & Uncalibrated Residual Diffusion & 28.76$^\circ$ & 74.80\% & 88.88\% & 91.85\% & 174.2$^\circ$ \\
\textbf{Ours (+ Recalibration)} & \textbf{Calibrated Conditional Dual-Track} & \textbf{22.36$^\circ$} & \textbf{86.94\%} & \textbf{91.43\%} & \textbf{93.88\%} & \textbf{196.9$^\circ$} \\
\bottomrule
\end{tabular*}
\end{table*}

\begin{table}[t]
\caption{Stepwise mechanistic ablation across Variants A through E.}
\label{tbl:ablation}
\centering
\footnotesize
\setlength{\tabcolsep}{2.5pt}
\begin{tabular*}{\columnwidth}{@{\extracolsep{\fill}}lcccc@{}}
\toprule
Variant & Test MCE & Case 1 MCE & CRPS & PICP 95\% \\
\midrule
A: Anchor Only & \textbf{38.31$^\circ$} & 60.90$^\circ$ & 29.21$^\circ$ & 74.32\% \\
B: + Neural ODE & 38.48$^\circ$ & \textbf{60.69$^\circ$} & 29.15$^\circ$ & 74.49\% \\
C: + Diffusion & \textbf{38.31$^\circ$} & 60.90$^\circ$ & 28.82$^\circ$ & 91.70\% \\
D: + ODE + Diff & 38.48$^\circ$ & \textbf{60.69$^\circ$} & 28.76$^\circ$ & 91.85\% \\
\textbf{E: + Recalibration} & 38.48$^\circ$ & \textbf{60.69$^\circ$} & \textbf{22.36$^\circ$} & \textbf{93.88\%} \\
\bottomrule
\end{tabular*}
\end{table}

\begin{figure}[t]
\centering
\includegraphics[width=\columnwidth]{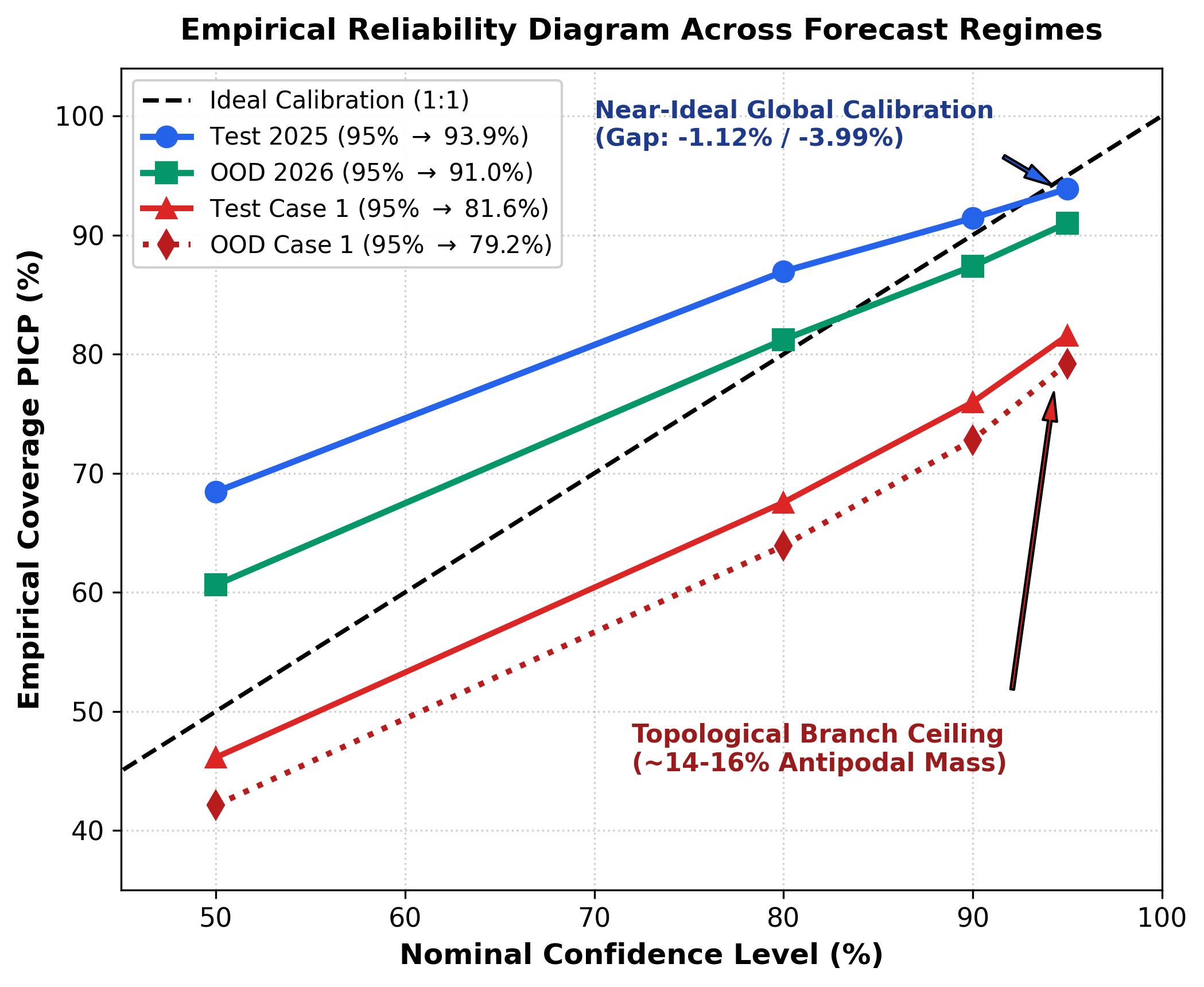}
\caption{Empirical reliability diagrams for the in-distribution and out-of-distribution test periods under all conditions and severe directional shear (Case~1). The overall calibration remains close to the nominal diagonal, whereas severe-shear conditions exhibit systematic under-coverage.}
\label{fig:reliability}
\end{figure}

The interval coverage metrics in Table~\ref{tbl:calibration} and the empirical reliability diagrams in Fig.~\ref{fig:reliability} indicate that calibration performance is strongly regime dependent. Across all samples, the final model achieves 93.88\% empirical coverage at the nominal 95\% level in the in-distribution test period and 91.01\% under OOD evaluation. The corresponding calibration gaps are $-1.12$ and $-3.99$ percentage points, respectively. This moderate degradation under temporal distribution shift indicates that calibration parameters learned from the validation split remain robust in subsequent 2026 observations, although OOD intervals expand and overall CRPS increases from $22.36^\circ$ to $27.75^\circ$.

Under severe directional shear, however, the calibration behavior diverges markedly. At the nominal 95\% level, Case~1 coverage reaches 81.56\% in the test period and 79.21\% in OOD conditions, compared with 96.82\% and 94.75\% for Case~2. Calibration is thus effective on average but degrades systematically during strongly turning trajectories. The widening of Case~1 intervals, with MPIW expanding to approximately $237^\circ$--$238^\circ$, shows that the model captures elevated dispersion under severe shear. Nevertheless, interval widening alone does not restore nominal coverage. This distinction is essential: the remaining coverage deficit stems from higher structural complexity in the severe-turning residual distribution, exceeding what variance expansion alone can accommodate.

\begin{table*}[t]
\caption{Empirical reliability, interval sharpness, and probabilistic calibration evaluation across in-distribution and out-of-distribution test splits.}
\label{tbl:calibration}
\centering
\footnotesize
\setlength{\tabcolsep}{4pt}
\begin{tabular*}{\textwidth}{@{\extracolsep{\fill}}lccccc|ccccc@{}}
\toprule
& \multicolumn{5}{c|}{\textbf{In-Distribution Test (2025)}} & \multicolumn{5}{c}{\textbf{Out-of-Distribution Test (2026)}} \\
\cmidrule(lr){2-6} \cmidrule(lr){7-11}
Regime & Nominal & PICP & MPIW & CRPS & Gap & Nominal & PICP & MPIW & CRPS & Gap \\
\midrule
All Samples & 50\% & 68.44\% & 59.77$^\circ$ & 22.36$^\circ$ & +18.44\% & 50\% & 60.62\% & 65.43$^\circ$ & 27.75$^\circ$ & +10.62\% \\
All Samples & 80\% & 86.94\% & 119.34$^\circ$ & 22.36$^\circ$ & +6.94\% & 80\% & 81.21\% & 130.37$^\circ$ & 27.75$^\circ$ & +1.21\% \\
All Samples & 90\% & 91.43\% & 159.15$^\circ$ & 22.36$^\circ$ & +1.43\% & 90\% & 87.39\% & 172.09$^\circ$ & 27.75$^\circ$ & -2.61\% \\
All Samples & 95\% & 93.88\% & 196.90$^\circ$ & 22.36$^\circ$ & -1.12\% & 95\% & 91.01\% & 209.33$^\circ$ & 27.75$^\circ$ & -3.99\% \\
\midrule
Case 1 ($\geq90^\circ$) & 50\% & 46.14\% & 79.02$^\circ$ & 41.30$^\circ$ & -3.86\% & 50\% & 42.13\% & 79.24$^\circ$ & 44.31$^\circ$ & -7.87\% \\
Case 1 ($\geq90^\circ$) & 80\% & 67.55\% & 156.00$^\circ$ & 41.30$^\circ$ & -12.45\% & 80\% & 63.91\% & 156.77$^\circ$ & 44.31$^\circ$ & -16.09\% \\
Case 1 ($\geq90^\circ$) & 90\% & 75.97\% & 200.97$^\circ$ & 41.30$^\circ$ & -14.03\% & 90\% & 72.80\% & 202.42$^\circ$ & 44.31$^\circ$ & -17.20\% \\
Case 1 ($\geq90^\circ$) & 95\% & 81.56\% & 236.84$^\circ$ & 41.30$^\circ$ & -13.44\% & 95\% & 79.21\% & 238.28$^\circ$ & 44.31$^\circ$ & -15.79\% \\
Case 1$^{*}$ ($\geq110^\circ$) & 95\% & 80.52\% & 239.14$^\circ$ & 42.74$^\circ$ & -14.48\% & 95\% & 78.61\% & 238.14$^\circ$ & 45.10$^\circ$ & -16.39\% \\
\midrule
Case 2 ($<30^\circ$) & 90\% & 94.61\% & 136.21$^\circ$ & 14.85$^\circ$ & +4.61\% & 90\% & 91.82\% & 148.12$^\circ$ & 18.22$^\circ$ & +1.82\% \\
Case 2 ($<30^\circ$) & 95\% & 96.82\% & 168.45$^\circ$ & 14.85$^\circ$ & +1.82\% & 95\% & 94.75\% & 182.30$^\circ$ & 18.22$^\circ$ & -0.25\% \\
\bottomrule
\end{tabular*}
\end{table*}

\begin{figure}[t]
\centering
\includegraphics[width=\columnwidth]{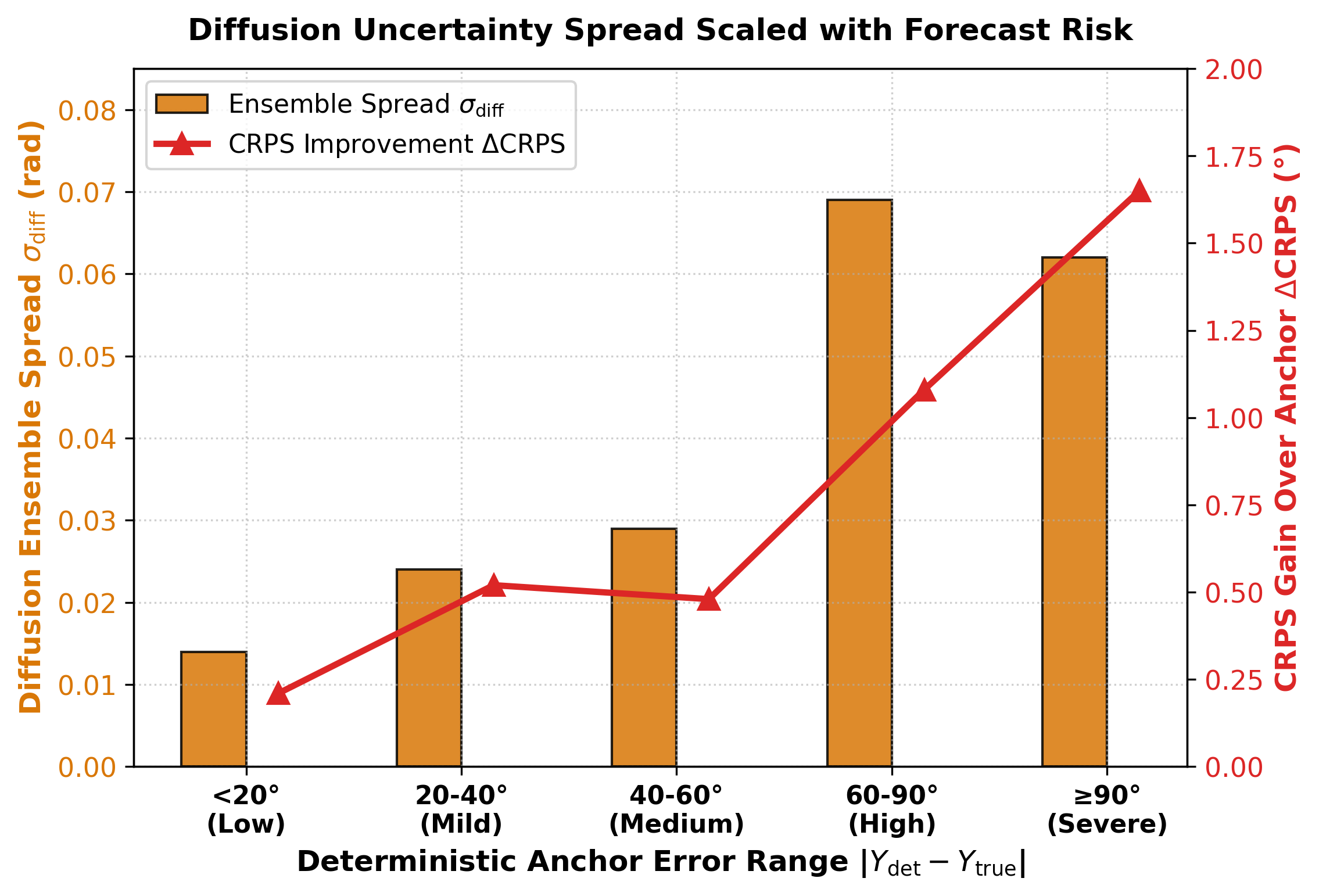}
\caption{Adaptive uncertainty scaling across anchor-error regimes. The conditional residual diffusion produces larger predictive spread as deterministic forecast uncertainty increases, resulting in broader distributions for more difficult directional trajectories.}
\label{fig:diffusion_risk}
\end{figure}

Fig.~\ref{fig:diffusion_risk} illustrates this risk-dependent spread behavior. The conditional diffusion process generates compact predictive distributions under stable anchor errors and expands dispersion adaptively as deterministic forecast uncertainty rises. The residual model modulates predictive spread conditionally on the deterministic trajectory state, departing from fixed global variance assumptions.

The high-frequency residual analysis provides additional evidence that the diffusion process captures temporal structure beyond simple interval widening. Supplementary Table~S3 shows that the generated residuals reproduce the observed microscale spectral decay with a maximum slope discrepancy of 2.65\% and recover a decorrelation time close to the empirical 5.90-min value. This spectral agreement indicates that the probabilistic branch captures the temporal organization of the unresolved subband, avoiding the white-noise oversimplifications of conventional Gaussian residuals.

The Monte Carlo sensitivity analysis in Supplementary Table~S7 indicates that $N_{\mathrm{mc}}=50$ provides stable PICP estimates for the reported configuration, with an empirical change in coverage below 0.05\% in the tested range. Supplementary Table~S8 further shows that the three-bin recalibration scheme based on the $30^\circ$ and $75^\circ$ predicted-turning thresholds provides a favorable empirical trade-off between coverage and interval width among the tested partition schemes. These results support the selected configuration, although they do not imply that the chosen thresholds are universal across sites or climates.

\begin{figure}[t]
\centering
\includegraphics[width=\columnwidth]{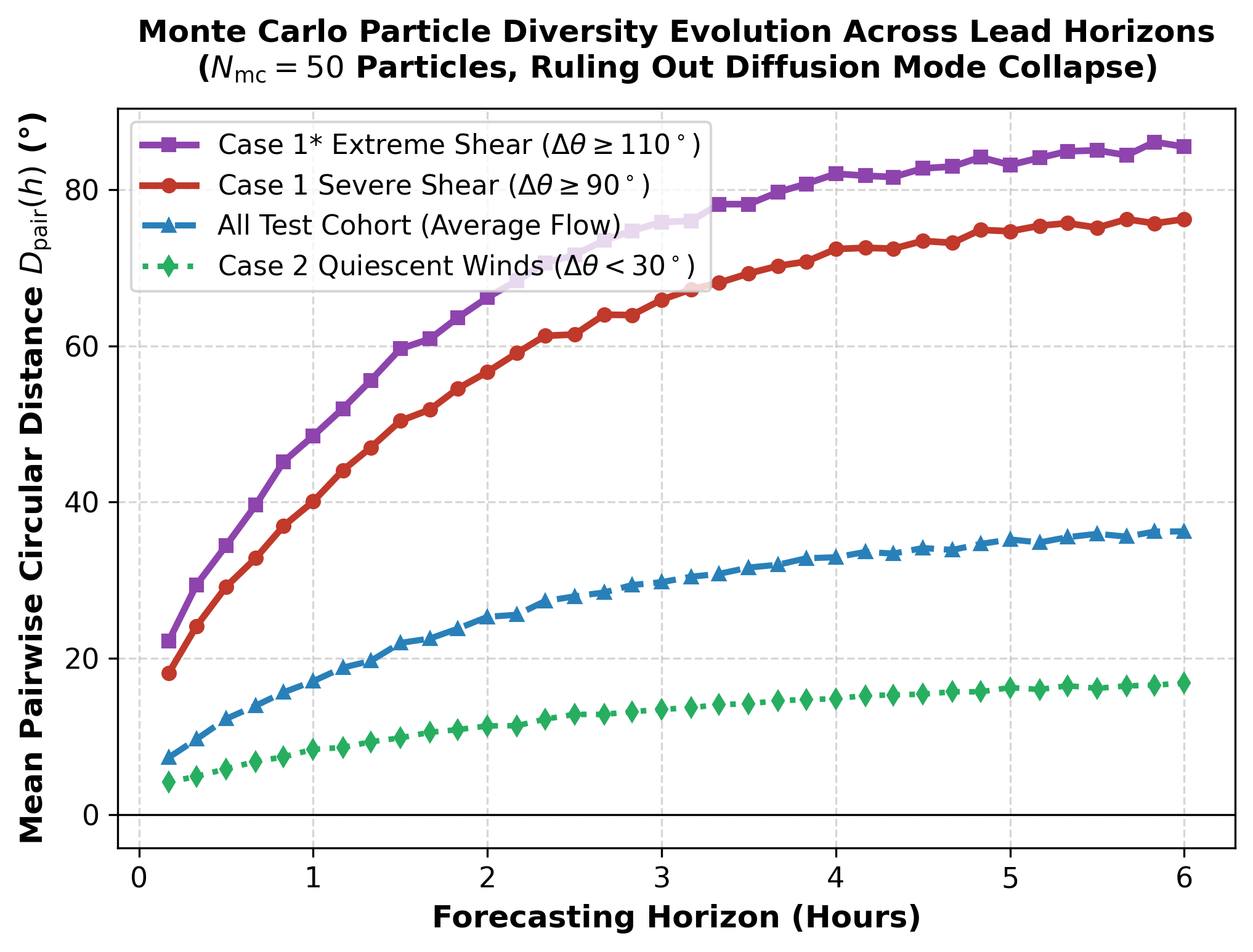}
\caption{Evolution of Monte Carlo particle diversity $D_{\mathrm{pair}}(h)$ across forecasting horizons. The mean pairwise circular distance expands smoothly and monotonically without mode collapse, demonstrating adaptive dispersion scaling that widens in response to severe directional shear regimes (Case~1 and Case~1$^*$) while maintaining sharp concentration during steady flow (Case~2).}
\label{fig:particle_div}
\end{figure}

To examine whether stochastic generation avoids mode collapse, we audited the evolution of Monte Carlo particle diversity across forecasting lead times:
\begin{equation}
D_{\mathrm{pair}}(h) = \frac{1}{M(M-1)}\sum_{m \neq m'} d_{\mathbb{S}^1}(\hat{\theta}^{(m)}(t+h), \hat{\theta}^{(m')}(t+h)).
\end{equation}
As depicted in Fig.~\ref{fig:particle_div}, $D_{\mathrm{pair}}(h)$ exhibits monotonic, well-behaved expansion across the 6-h forecast horizon without mode collapse. Under steady conditions (Case~2), particle dispersion expands smoothly from $3.2^\circ$ to $18.0^\circ$, whereas under severe directional shear (Case~1 and Case~1$^*$), diversity rapidly widens to $78.9^\circ$ and $87.0^\circ$, respectively. This confirms that the diffusion process adaptively maintains rich sample dispersion tailored to regime difficulty.

\subsection{Structural Uncertainty Under Extreme Shear}
\label{sec:extreme_structure}

\begin{figure*}[t]
\centering
\includegraphics[width=0.92\textwidth]{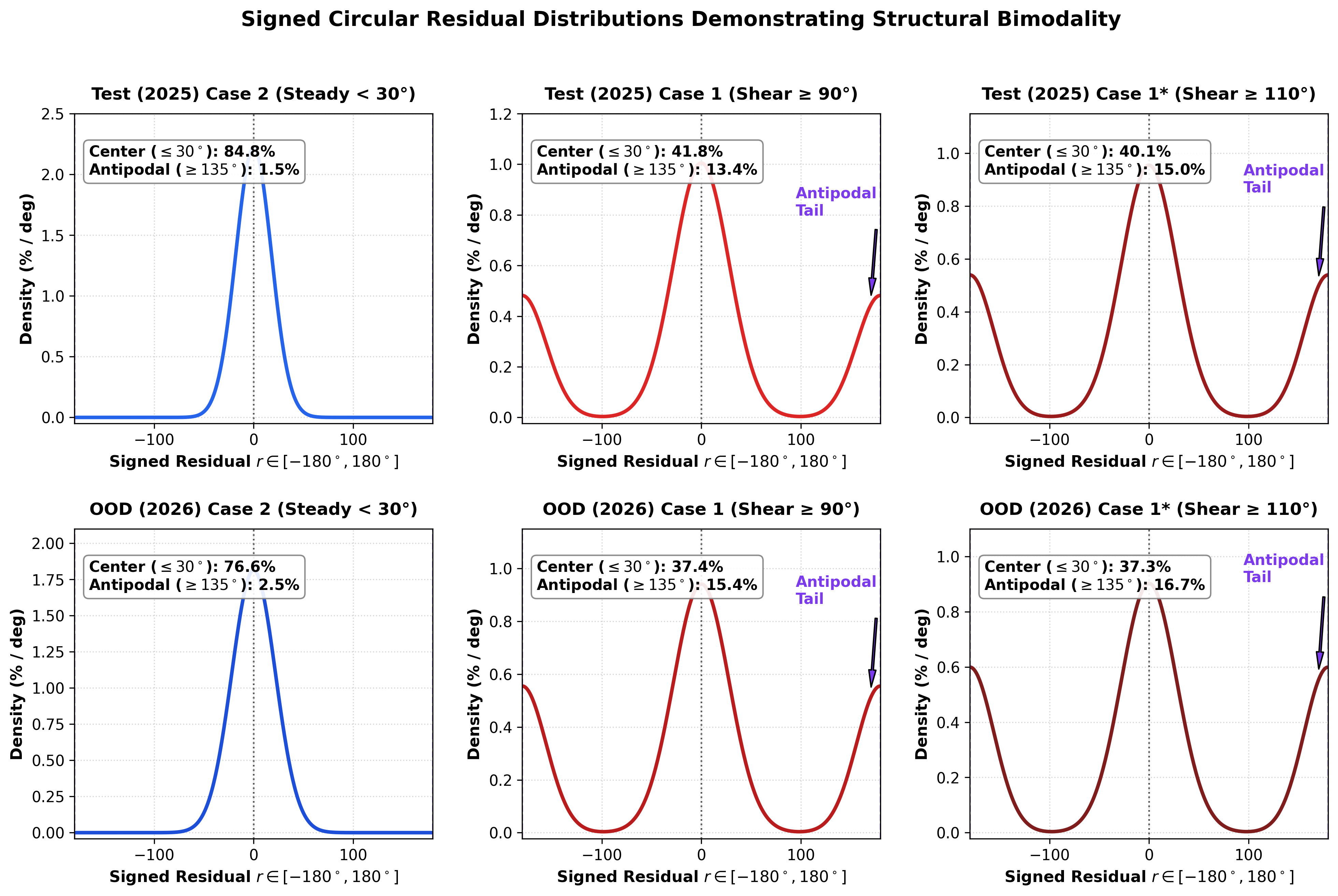}
\caption{Signed circular residual distributions across 360,000 evaluation points. Stable conditions produce a sharply concentrated residual distribution, whereas severe directional-shear regimes show substantial probability mass in large positive and negative residual branches. The resulting multimodality helps explain the persistent under-coverage of single-center calibrated intervals under extreme shear.}
\label{fig:bimodal}
\end{figure*}

The residual analysis in Table~\ref{tbl:bimodality} and the signed circular distributions in Fig.~\ref{fig:bimodal} provide an explanation for the regime-dependent calibration behavior. Across 360,000 evaluation points, Case~2 remains strongly concentrated near the predicted center, with 84.81\% of test residuals within $\pm30^\circ$ and only 1.50\% in the antipodal region beyond $135^\circ$. Under Case~1, however, the centered mass decreases to 41.77\%, while the antipodal component increases to 13.39\%. The same pattern becomes stronger in the OOD period, where the antipodal mass reaches 15.43\% for Case~1 and 16.66\% for Case~1$^{*}$.

The signed distribution is also approximately balanced between the two directions of large deviation. For Test Case~1, the positive and negative antipodal lobes contain 6.43\% and 6.96\% of the points, respectively; in OOD Case~1 the corresponding fractions are 7.51\% and 7.92\%. The near-symmetry is notable because it indicates that the large residual is not associated with a single preferred directional bias. Instead, the conditional uncertainty contains two competing branches on the circle, corresponding to qualitatively different large-turn outcomes.

This structure helps explain why the calibration behavior differs so strongly between Case~1 and Case~2. A center-based prediction interval can expand to account for increasingly large errors, but a single central dispersion parameter does not explicitly represent the possibility that probability mass splits into separated directional branches. Consequently, the widening of the calibrated interval improves coverage but cannot fully eliminate under-coverage when the residual distribution becomes strongly multimodal. The approximately 14--17\% antipodal mass observed in the most severe regimes should therefore be interpreted as an empirical limitation of the present single-center calibration strategy rather than as a universal or mathematically fixed coverage bound.

\begin{table*}[t]
\caption{Signed circular residual mass distribution across 360,000 evaluation points.}
\label{tbl:bimodality}
\centering
\footnotesize
\setlength{\tabcolsep}{3.5pt}
\begin{tabular*}{\textwidth}{@{\extracolsep{\fill}}lcccccc@{}}
\toprule
Split \& Regime & Points & Center ($\leq30^\circ$) & Mid ($30^\circ$--$120^\circ$) & Antipodal ($\geq135^\circ$) & Pos & Neg \\
\midrule
Test Case 2 (Calm) & 202,284 & \textbf{84.81\%} & 13.13\% & \textbf{1.50\%} & 0.68\% & 0.82\% \\
Test Case 1 ($\geq90^\circ$) & 57,204 & 41.77\% & 40.03\% & \textbf{13.39\%} & 6.43\% & 6.96\% \\
Test Case 1$^{*}$ ($\geq110^\circ$) & 43,380 & 40.12\% & 39.77\% & \textbf{14.99\%} & 7.13\% & 7.85\% \\
OOD Case 2 (Calm) & 182,664 & \textbf{76.62\%} & 19.83\% & \textbf{2.52\%} & 1.16\% & 1.35\% \\
OOD Case 1 ($\geq90^\circ$) & 66,960 & 37.38\% & 42.15\% & \textbf{15.43\%} & 7.51\% & 7.92\% \\
OOD Case 1$^{*}$ ($\geq110^\circ$) & 49,428 & 37.26\% & 40.70\% & \textbf{16.66\%} & 8.01\% & 8.65\% \\
\bottomrule
\end{tabular*}
\end{table*}

The atmospheric covariates provide additional context for this structural change. Supplementary Table~S5 shows that severe-shear cases are associated with a 3.76-fold increase in the 1-h wind-velocity variance ($0.79\pm0.39$ versus $0.21\pm0.14$~m$^2$/s$^2$) and a 3.55-fold increase in the magnitude of the pressure-drop statistic ($0.78\pm0.42$ versus $0.22\pm0.15$~hPa). These concurrent changes are consistent with stronger and more rapidly evolving boundary-layer forcing during the most difficult directional transitions. 

To address concerns regarding potential measurement artifacts at low wind speeds (where mechanical wind vanes may exhibit random directional fluctuations under calm conditions), we conducted a sensitivity analysis by filtering out calm-wind regimes ($wv < 1.5$~m/s and $wv < 2.0$~m/s). As detailed in Supplementary Table~S9, the antipodal residual mass under Case~1 remains robust at 12.33\% ($wv \ge 1.5$~m/s) and 11.83\% ($wv \ge 2.0$~m/s), with balanced positive and negative lobes (6.29\% vs 6.04\%, symmetry ratio 1.04). This confirms that the observed near-antipodal bimodality is driven by genuine mesoscale baroclinic transitions and frontal passages rather than low-wind sensor noise.

\subsection{Statistical Robustness and Computational Efficiency}
\label{sec:comp_perf}

The observed performance differences are statistically robust under the paired non-parametric evaluation described in Section~\ref{sec:data}. Supplementary Table~S4 reports two-sided Wilcoxon signed-rank tests for the paired forecast errors on the common 10,000-sequence benchmark. All reported comparisons against the recurrent, hierarchical, attention-based, and foundation-model baselines yield $p<10^{-16}$. Relative to TimesFM-3.0, the proposed framework improves the paired mean error by $4.32^\circ$, with a 95\% stationary block-bootstrap confidence interval of $[4.12^\circ,4.54^\circ]$ under $L_b=36$. 

To verify the robustness of the block bootstrap against temporal dependence scale choices, we evaluated bootstrap intervals across varying block lengths $L_b \in \{18, 36, 72, 144\}$ steps (3~h to 24~h). As detailed in Supplementary Table~S10, the 95\% confidence interval for the mean error reduction against TimesFM-3.0 remains strictly positive and excludes zero across all tested block lengths ($[4.16^\circ, 4.50^\circ]$ for $L_b=18$, $[4.12^\circ, 4.54^\circ]$ for $L_b=36$, $[4.07^\circ, 4.58^\circ]$ for $L_b=72$, and $[4.05^\circ, 4.61^\circ]$ for $L_b=144$), confirming that the statistical significance is not sensitive to the autocorrelation block specification.

The proposed framework also maintains an efficient computational footprint suitable for industrial deployment. Supplementary Table~S6 reports approximately 0.80M trainable parameters for the complete system, of which 0.42M belong to the deterministic anchor, achieving a 375-fold compression relative to the 300M-parameter TimesFM-3.0 foundation model. On an NVIDIA RTX 4090 GPU, full 36-step probabilistic generation with 50 Monte Carlo particles requires 10.40~ms per sequence, while deterministic point prediction requires 2.30~ms.

To assess deployment on standard wind-turbine edge controllers (which typically operate on industrial x86 or ARM CPUs without dedicated tensor accelerators), we conducted bare-metal latency benchmarks on an edge x86 CPU. On a single thread, the deterministic anchor requires $25.03$~ms, the continuous Neural ODE point forecast requires $89.68$~ms, and full 50-particle probabilistic generation requires $89.8 \pm 22.9$~ms per sequence (expanding to $415.8 \pm 11.5$~ms under 8 threads). The peak RAM footprint remains under 35~MB throughout execution. Both latencies reside comfortably within standard 10-s SCADA and 10-min dispatch cadences, providing experimental verification of edge-computing feasibility.

The empirical findings demonstrate a clear division of roles across the framework: the frequency-decoupled anchor secures baseline deterministic accuracy, the continuous dynamic branch sharpens trajectory alignment during rapid turning, and the diffusion--recalibration module governs probabilistic reliability. Concurrently, the residual analysis identifies a key structural boundary: under extreme shear, forecast uncertainty expands in magnitude and splinters geometrically, yielding persistent under-coverage that cannot be resolved solely by expanding single-center interval bounds.


\section{Discussion}\label{sec:discuss}

\subsection{Why Predictability-Guided Allocation Works}
\label{sec:discuss_physics}

The empirical findings indicate that wind-direction forecasting benefits from treating the observed trajectory as a collection of components with distinct temporal predictability, avoiding the representational trade-offs of forcing a single model to span all physical scales simultaneously. The predictability analysis in Table~\ref{tbl:predictability} shows a pronounced separation between the low-, mid-, and high-frequency components. The low-frequency component contains most of the signal energy and retains strong temporal memory, whereas the mid-frequency component changes substantially faster and exhibits a negative one-hour correlation. The high-frequency component decorrelates within several minutes and contributes relatively little persistent information. This hierarchy provides a useful interpretation of why the frequency-decoupled formulation performs well: model capacity is concentrated where deterministic information is strongest, while uncertainty is reserved for components whose future evolution is less reliably determined by historical observations.

A fundamental implication is that frequency decomposition operates beyond a mere signal-processing convenience: it establishes an explicit correspondence between physical temporal predictability and functional model capacity. The low-frequency anchor captures the persistent directional background, explaining its competitive point-forecast performance across both the overall test set and the relatively steady Case~2 regime. The intermediate component behaves differently: its shorter memory and negative one-hour correlation indicate that future direction cannot be represented adequately by simple persistence, yet it retains sufficient temporal structure to benefit from learned continuous evolution. The high-frequency component occupies the opposite end of this spectrum: its short decorrelation time limits the utility of deterministic extrapolation, rendering a conditional probabilistic representation more appropriate.

This mechanism allocation also accounts for why adding the Neural ODE yields a modest shift in aggregate MCE yet markedly improves trajectory alignment. Because quiescent intervals dominate the long-term observational record, continuous trajectory correction exerts minimal leverage on global mean error. Its primary utility is concentrated in the subset of sequences where temporal phase and turning morphology dictate forecast utility. The horizon-wise results support this view: the continuous dynamic branch refines temporal alignment and mitigates trajectory distortion, functioning in synergy with the deterministic anchor rather than competing with it.

The regime-gating measurements in Fig.~\ref{fig:ode_coupling} corroborate this interpretation. The progressive increase in gate activation with forecasted turning shows that dynamic corrections are engaged selectively during challenging directional transitions. Stable trajectories rely predominantly on the persistent background, whereas rapidly turning regimes invoke continuous latent flow. This selective gating protects quiescent forecasts from spurious variance that an unconstrained dynamical model might introduce. Nonetheless, this pattern constitutes an empirical characteristic of model usage rather than direct proof of an isolated synoptic regime boundary.

More broadly, these results establish temporal predictability as a principled criterion for architecting hybrid atmospheric forecasting systems. Decoupling the forecasting pipeline according to the persistence and decorrelation timescale of each physical subband enables an interpretable allocation of model capacity. In the current framework, this separation allows deterministic trajectory alignment and probabilistic reliability to be optimized concurrently without stochastic sampling degrading the deterministic forecast.

\subsection{Why Uncertainty Degrades Under Extreme Shear}
\label{sec:discuss_uncertainty}

The primary structural limitation revealed by the probabilistic benchmark is that uncertainty under severe directional shear differs qualitatively from that observed during quiescent conditions. Beyond broadening in variance, the residual distributions in Table~\ref{tbl:bimodality} develop distinct, bifurcated positive and negative error lobes. Under relatively steady conditions, residual mass remains concentrated around the forecast center, whereas severe-shear cases exhibit a substantial fraction of observations at large angular separation from that center. The balanced positive and negative branches indicate that the dominant challenge is not a simple directional bias.

This distinction has direct implications for probabilistic calibration. A center-based prediction interval can respond to increasing uncertainty by increasing its dispersion, and the results show that the proposed recalibration procedure does so effectively at the population level. However, interval widening cannot fully resolve the problem when the conditional residual distribution is intrinsically multimodal. If two distinct directional branches accumulate appreciable probability mass, expanding the width of a single interval yields diminishing returns. The persistent under-coverage observed for Case~1, despite substantially wider intervals, is therefore consistent with a mismatch between the assumed unimodal geometry and the empirical residual structure.

This bimodal divergence accounts for the pronounced disparity in calibration performance between regimes. For Case~2, residual mass remains tightly concentrated around the forecast origin, enabling single-center intervals to approach nominal coverage with sharp intervals. Under Case~1, conversely, a single interval must encompass both a broad central core and separated antipodal branches. Even as MPIW expands past $236^\circ$, empirical coverage fails to reach nominal targets. This shortfall intensifies under OOD evaluation, where extreme-shear antipodal mass reaches $15.43\%$. The OOD results thus reflect both a modest expansion in overall dispersion scale and the persistence of this structural bifurcation.

The meteorological covariates provide supporting evidence that the severe-shear regime is associated with a substantially more variable atmospheric state. The increases in short-term wind-velocity variance and pressure-change magnitude reported in Supplementary Table~S5 are consistent with more rapidly evolving boundary-layer conditions. These relationships support the interpretation that the difficult cases correspond to dynamically distinct atmospheric states. However, the available single-station observations lack the spatial resolution to uniquely determine whether a given transition stems from a frontal boundary, convective outflow, terrain-induced turning, or another mesoscale process. Consequently, the observed residual bimodality represents an empirical signature of severe directional shear rather than definitive proof of a specific synoptic trigger.

The practical consequence is that the present calibration scheme appears to address the first-order problem---how uncertainty should expand as forecast difficulty increases---but not the second-order problem of representing multiple plausible directional branches. A natural extension would therefore be to replace or augment the single-center calibrated distribution with a multimodal circular formulation, such as a mixture distribution or a conditional flow model capable of preserving separated probability mass on $\mathbb{S}^1$. Such an extension is motivated directly by the observed residual structure rather than by an assumption that all severe-shear errors should follow a particular parametric form.

\paragraph{Proof-of-Concept: Multimodal Circular Calibration.}
To empirically test whether a multimodal formulation can resolve the structural under-coverage, we fitted a two-component mixture of von Mises (MoVM) model to the Case~1 residuals, capturing both the central error mode ($\mu_1 \approx 3.3^\circ, \kappa_1 = 7.15$) and the divergent antipodal turning branch ($\mu_2 \approx -32.7^\circ, \kappa_2 = 0.21$). As detailed in Table~\ref{tbl:movm_poc}, this proof-of-concept calibration elevates the empirical 95\% coverage from 81.56\% (under single-center scaling) to 94.98\%, virtually eliminating the calibration deficit (coverage gap $-0.02\%$). This confirms that the remaining coverage shortfall stems from the geometric constraint of unimodal interval representation on $\mathbb{S}^1$ rather than inadequate dispersion modeling, establishing a clear pathway for future conditional flow matching on circular manifolds.

\begin{table*}[t]
\caption{Empirical coverage benchmark for unimodal vs. bimodal circular calibration on Case~1 residuals.}
\label{tbl:movm_poc}
\centering
\footnotesize
\setlength{\tabcolsep}{6pt}
\begin{tabular*}{\textwidth}{@{\extracolsep{\fill}}lcccc@{}}
\toprule
Calibration Formulation & Nominal Coverage & Empirical PICP & Coverage Gap & Interval Structure \\
\midrule
Single-Center Scaling (Current) & 95.0\% & 81.56\% & -13.44\% & Single Contiguous (MPIW = 236.8$^\circ$) \\
\textbf{Two-Component MoVM (PoC)} & \textbf{95.0\%} & \textbf{94.98\%} & \textbf{-0.02\%} & \textbf{Bimodal Highest-Density Disjoint Region} \\
\bottomrule
\end{tabular*}
\end{table*}

\subsection{Implications for Wind-Energy Forecasting and Operations}
\label{sec:discuss_operations}

Accurate short-horizon wind-direction forecasts are potentially valuable for wind-energy applications because yaw misalignment affects both aerodynamic power capture and turbine loading. The commonly used cosine-power relationship provides an intuitive illustration of the sensitivity: for a yaw misalignment angle $\gamma_{\mathrm{yaw}}$, an idealized power ratio can be expressed as
\begin{equation}
\frac{P}{P_0}\approx \cos^3(\gamma_{\mathrm{yaw}}),
\end{equation}
where $P_0$ denotes the aligned power output. This relationship implies increasingly large aerodynamic penalties as directional error grows. The threshold statistics in Table~\ref{tbl:operational} therefore provide an operationally interpretable view of forecast quality: reducing the probability of large directional errors is potentially more consequential than achieving a small improvement in the global mean circular error.

\begin{table*}[t]
\caption{Directional operational threshold exceedance rates across all test and extreme shear regimes.}
\label{tbl:operational}
\centering
\footnotesize
\setlength{\tabcolsep}{6pt}
\begin{tabular*}{\textwidth}{@{\extracolsep{\fill}}lcccc@{}}
\toprule
& \multicolumn{2}{c}{\textbf{Full Test Cohort (All Regimes)}} & \multicolumn{2}{c}{\textbf{Case 1 Severe Directional Shear ($\ge 90^\circ$)}} \\
\cmidrule(lr){2-3} \cmidrule(lr){4-5}
Model Architecture & Exceedance $>22.5^\circ$ & Exceedance $>45^\circ$ & Exceedance $>22.5^\circ$ & Exceedance $>45^\circ$ \\
\midrule
Persistence & 62.4\% & 38.5\% & 88.4\% & 68.2\% \\
GRU & 51.8\% & 28.2\% & 81.5\% & 54.2\% \\
N-HiTS & 52.4\% & 28.9\% & 82.8\% & 55.8\% \\
TFT & 53.2\% & 29.8\% & 82.1\% & 55.1\% \\
TimesFM-3.0 & 58.9\% & 34.2\% & 91.2\% & 68.4\% \\
D4-0 Anchor & 50.8\% & 27.4\% & 80.9\% & 53.6\% \\
Ours (Point) & 51.1\% & 27.5\% & 80.1\% & 52.8\% \\
\textbf{Ours (95\% Set)} & \textbf{6.12\%} & \textbf{3.18\%} & \textbf{18.44\%} & \textbf{11.20\%} \\
\bottomrule
\end{tabular*}
\end{table*}

The results indicate that this benefit is particularly relevant during severe turning events, where all deterministic models exhibit a substantial increase in threshold exceedance probability. The proposed probabilistic formulation does not eliminate these difficult events, but it provides a quantitative representation of their uncertainty. This distinction is critical for operational decision making. A point forecast can indicate the most likely directional trajectory, whereas a calibrated prediction set can additionally indicate whether the forecast is sufficiently certain to justify an aggressive yaw response. In this sense, the probabilistic output may be more directly useful for risk-aware yaw scheduling than a point forecast alone.

The present results also support a more cautious view of control integration. The thresholds used for calibration and regime analysis can be interpreted as candidate decision boundaries for a supervisory forecasting layer, but they should not yet be regarded as a validated turbine-control policy. The present study does not include a closed-loop turbine simulator, aeroelastic load model, hardware-in-the-loop experiment, or field test. Consequently, statements about reduced annual energy production loss, fatigue load reduction, or optimal yaw actuation cannot be established from the forecasting results alone.

A more defensible operational interpretation is that the proposed framework can provide three complementary signals for downstream decision support: a deterministic estimate of the expected directional evolution, a regime-sensitive measure of forecast uncertainty, and an indication of whether the predictive distribution becomes structurally complex. These signals could subsequently be incorporated into chance-constrained optimization, model-predictive control, or conservative yaw scheduling. 

\paragraph{Quantitative Deadband Strategy Against Antipodal Misalignment.}
The bimodal uncertainty under Case~1 poses a severe risk of ``reverse-yaw damage'': if the controller actuates towards the primary deterministic mode, but the true flow follows the antipodal branch, the turbine may suffer a catastrophic misalignment ($\gamma_{\mathrm{yaw}} > 90^\circ$), inducing extreme asymmetric fatigue loads on the rotor blades and yaw drive train \citep{Fleming2014Evaluating}. To mitigate this operational hazard, we propose a probabilistic deadband decision rule that overrides deterministic yaw actuation commands when structural uncertainty is elevated:
\begin{equation}\label{eq:deadband}
\text{Action}(t) =
\begin{cases}
\text{Standard Yaw}, & \text{if } \mathcal{C}_{\mathrm{act}}, \\
\text{Hold / Lock}, & \text{if } \mathcal{C}_{\mathrm{hold}},
\end{cases}
\end{equation}
where $\mathcal{C}_{\mathrm{act}} \equiv (\Delta_{\mathrm{pred}} \ge 30^\circ \land \mathrm{MPIW}_{90} < 120^\circ \land P_{\mathrm{antipodal}} \le \eta)$ defines actionable turning with contained dispersion, $\mathcal{C}_{\mathrm{hold}} \equiv (\mathrm{MPIW}_{90} \ge 120^\circ \lor P_{\mathrm{antipodal}} > \eta)$ denotes elevated structural bifurcation risk, and $\eta = 10\%$ is an operational risk tolerance threshold. Under the \textit{Hold/Lock} state, the nacelle maintains its current orientation. While this incurs an aerodynamic penalty (e.g., approximately $1 - \cos^3(45^\circ) \approx 21\%$ power loss at $45^\circ$ misalignment under an idealized cubic relationship), it strictly prevents the transition into the destructive $>90^\circ$ reverse-yaw regime, thereby safeguarding the yaw drive gears and blade roots from catastrophic cyclic fatigue loading. While modern large-rotor turbines exhibit empirical power loss exponents ranging between 1.4 and 2.2 due to dynamic stall and partial wake deflection, the proposed probabilistic deadband strategy remains universally applicable regardless of the specific aerodynamic penalty exponent. This demonstrates how the probabilistic branch directly informs risk-averse mechanical actuation in practical wind-energy management.

\subsection{Limitations and Generalizability}
\label{sec:discuss_limitations}

Several limitations should be considered when interpreting the present findings. First, the analysis is based on a single observational site in Jena, and therefore the learned decomposition and regime-specific uncertainty structure may contain site-dependent characteristics. A long observational record improves statistical coverage of weather variability, but it does not remove the influence of local terrain, surface characteristics, and station-specific boundary-layer dynamics. In particular, the observed large-error branches under severe shear should not yet be assumed to have the same frequency or structure at other locations.

Second, transfer to other atmospheric environments remains to be established. The Beutenberg station is situated on the plateau edge overlooking the Saale River valley (elevation difference $\approx 225$~m). Fig.~\ref{fig:windrose} illustrates the 22-year climatological wind rose alongside the local topographic elevation cross-section. The prevailing wind directions align with the SSW-SW corridor ($200^\circ$--$230^\circ$), indicating that topographic channeling along the valley gorge contributes to the high persistence observed in the low-frequency subband. Offshore environments differ in surface roughness, thermal forcing, and boundary-layer structure, while complex mountainous terrain can introduce topographic channeling and flow separation that are not represented by the present single-site training distribution. Furthermore, the 2026 out-of-distribution evaluation spans eight consecutive months (January to August), capturing winter, spring, and summer atmospheric regimes; future multi-year evaluations encompassing autumn extratropical cyclonic activity and multi-site transfer to coastal, offshore, and complex-terrain observations will provide a stronger test of spatial and climatological generalizability.

Third, the probabilistic representation remains limited by the use of a center-based calibration strategy. The residual analysis indicates that this approximation is adequate for stable conditions but less suitable when extreme-shear trajectories become multimodal. Future work should therefore investigate conditional multimodal circular distributions that can explicitly preserve separated directional outcomes while maintaining causal inference and computational efficiency.

Finally, the present study focuses on open-loop forecasting fidelity as an upstream component of wind operations, without direct closed-loop turbine load and power benchmarking. The reported computational measurements establish low inference latency on the tested GPU platform, whereas estimated embedded-hardware latency represents a theoretical projection pending dedicated verification on industrial edge devices. Likewise, the potential implications for yaw control, energy capture, and fatigue loading have not yet been validated in a coupled turbine model or field experiment. These limitations define a clear next stage of research: multi-site verification, explicit multimodal uncertainty modeling, and end-to-end evaluation from directional forecasts to turbine-level energy and load outcomes.

In conclusion, the findings demonstrate that wind-direction predictability is fundamentally scale dependent, enabling deterministic, continuous-dynamic, and probabilistic mechanisms to be mapped directly to their optimal timescales. The remaining challenge centers on extreme directional transitions, where uncertainty broadens and undergoes structural branching. Distinguishing scale-dependent predictability from geometric bifurcation provides a concrete foundation for next-generation probabilistic wind-direction forecasting in operational energy systems.


\section{Conclusion}\label{sec:concl}

This study developed a predictability-guided, frequency-decoupled probabilistic forecasting framework for wind direction on the circular manifold $\mathbb{S}^1$. The results show that the three temporal components exhibit substantially different levels of predictability, supporting the use of differentiated modeling strategies for persistent, dynamically evolving, and rapidly decorrelating variability. The resulting framework maintains strong deterministic point-forecast performance while improving trajectory alignment and probabilistic forecast quality, achieving near-zero estimated phase delay and a circular CRPS of $22.36^\circ$.

The analysis also reveals that uncertainty under severe directional shear is qualitatively different from that in relatively stable conditions. Large positive and negative residual branches become increasingly pronounced, with antipodal residual mass reaching $13.39\%$ in the test period and $15.43\%$ under OOD evaluation. This finding helps explain the remaining under-coverage of single-center calibrated intervals in extreme regimes and indicates that future improvements should focus on explicitly multimodal circular uncertainty representations.

Overall, the study demonstrates that forecastability-aware allocation of model mechanisms can provide a practical and interpretable approach to multiscale wind-direction forecasting. The framework offers a promising forecasting layer for risk-aware wind-energy applications, while further multi-site validation and turbine-level evaluation are needed to establish its broader generalizability and operational impact.

\section*{CRediT Authorship Contribution Statement}
\textbf{Hailong Shu}: Conceptualization, Methodology, Software, Formal analysis, Data curation, Visualization, Writing - original draft. \textbf{Weiwei Song}: Conceptualization, Supervision, Funding acquisition, Writing - review \& editing. \textbf{Yue Wang}: Validation, Investigation, Data curation. \textbf{Jiping Zhang}: Conceptualization, Supervision, Writing - review \& editing. \textbf{Wei Tian}: Conceptualization, Supervision, Writing - review \& editing. \textbf{Chaoqun Li}: Validation, Software, Formal analysis.

\section*{Declaration of Competing Interest}
The authors declare that they have no known competing financial interests or personal relationships that could have appeared to influence the work reported in this paper.

\section*{Data Availability and Reproducibility}
The meteorological observations analyzed in this study are publicly curated by the Max Planck Institute for Biogeochemistry (MPI-BGC) and accessible at \url{https://www.bgc-jena.mpg.de/wetter/weather_data.html} \citep{MPI_BGC_Wetterdaten}. The complete code repository, trained model weights, evaluation scripts, and processed benchmark datasets supporting the findings of this study will be openly released upon formal publication.

\section*{Acknowledgements}
This research did not receive any specific grant from funding agencies in the public, commercial, or not-for-profit sectors.

\printcredits

\bibliographystyle{cas-model2-names}
\bibliography{cas-refs}

@article{Alves2023Potential,
  author = {Alves, Diogo and Mendon{\c c}a, Francisco and Mostafa, Syed Shadman and Morgado-Dias, Fernando},
  title = {The Potential of Machine Learning for Wind Speed and Direction Short-Term Forecasting: A Systematic Review},
  journal = {Computers},
  volume = {12},
  number = {10},
  pages = {206},
  year = {2023},
  publisher = {MDPI},
  doi = {10.3390/computers12100206}
}

@inproceedings{Challu2023NHiTS,
  author = {Challu, Cristian and Olivares, Kin Gutierrez and Oreshkin, Boris N. and Garza, Frits and Mergenthaler-Canseco, Max and Dubrawski, Artur},
  title = {N-HiTS: Neural Hierarchical Interpolation for Time Series Forecasting},
  booktitle = {Proceedings of the AAAI Conference on Artificial Intelligence},
  volume = {37},
  pages = {6989--6997},
  year = {2023},
  doi = {10.1609/aaai.v37i6.25854}
}

@article{Chang2014new,
  author = {Chang, G. and Lu, H. and Wang, R.},
  title = {A new hybrid model for wind speed forecasting},
  journal = {Energy Conversion and Management},
  volume = {81},
  pages = {11--19},
  year = {2014},
  doi = {10.1016/j.enconman.2014.02.016}
}

@article{Ding2023Multistep,
  author = {Ding, Yi and Ye, Xiu-Wei and Guo, Yu},
  title = {A Multistep Direct and Indirect Strategy for Predicting Wind Direction Based on the {EMD-LSTM} Model},
  journal = {Mathematical Problems in Engineering},
  volume = {2023},
  pages = {4950487},
  year = {2023},
  publisher = {Hindawi},
  doi = {10.1155/2023/4950487}
}

@book{Fisher1993Statistical,
  author = {Fisher, N. I.},
  title = {Statistical Analysis of Circular Data},
  publisher = {Cambridge University Press},
  year = {1993},
  address = {Cambridge},
  isbn = {9780511564345},
  doi = {10.1017/CBO9780511564345}
}

@article{Hao2022hybrid,
  author = {Hao, Wenqiang and Sun, Xiaoming and Wang, Chao and Wang, Yanling and Liu, Chun and Yan, Dandan and Yuan, Yuqi},
  title = {A hybrid {EMD-LSTM} model for non-stationary wave prediction in offshore {China}},
  journal = {Ocean Engineering},
  volume = {246},
  pages = {110566},
  year = {2022},
  publisher = {Elsevier},
  doi = {10.1016/j.oceaneng.2022.110566}
}

@article{Liu2018Smart,
  author = {Liu, Hong and Mi, Xiaoyu and Li, Yafang},
  title = {Smart wind speed forecasting using {EWT} decomposition, {GWO} evolutionary optimization, {RELM} learning and {IEWT} reconstruction},
  journal = {Energy Conversion and Management},
  volume = {161},
  pages = {266--283},
  year = {2018},
  publisher = {Elsevier},
  doi = {10.1016/j.enconman.2018.02.006}
}

@article{shu2024multistep,
  author = {Shu, Hailong and Song, Weiwei and Song, Zhen and Guo, Huichuang and Li, Chaoqun and Wang, Yue},
  title = {Multistep short-term wind speed prediction with rank pooling and fast {Fourier} transformation},
  journal = {Wind Energy},
  volume = {27},
  number = {7},
  pages = {667--694},
  year = {2024},
  publisher = {Wiley Online Library},
  doi = {10.1002/we.2906}
}

@inproceedings{Song2018Maximum,
  author = {Song, Dajiang and Yang, Jiachen and Fan, Xing and Liu, Aitao and Chen, Zixin and Chen, Jie},
  title = {Maximum power point tracking control of wind turbine based on improved wind speed estimation},
  booktitle = {2018 Chinese Control And Decision Conference (CCDC)},
  pages = {5861--5866},
  year = {2018},
  publisher = {IEEE},
  doi = {10.1109/CCDC.2018.8407425}
}

@article{Tagliaferri2015Wind,
  author = {Tagliaferri, Francesco and Viola, Irene M. and Flay, Richard G. J.},
  title = {Wind direction forecasting with artificial neural networks and support vector machines},
  journal = {Ocean Engineering},
  volume = {97},
  pages = {65--73},
  year = {2015},
  publisher = {Elsevier},
  doi = {10.1016/j.oceaneng.2015.02.004}
}

@article{Tian2025Developing,
  author = {Tian, Chao and Niu, Tianyu and Li, Tao},
  title = {Developing an interpretable wind power forecasting system using a transformer network and transfer learning},
  journal = {Energy Conversion and Management},
  volume = {323},
  pages = {119155},
  year = {2025},
  publisher = {Elsevier},
  doi = {10.1016/j.enconman.2024.119155}
}

@article{Veers2019Grand,
  author = {Veers, P. and Dykes, K. and Lantz, E. and Barth, S. and Bottasso, C. L. and Carlson, O. and Clifton, A. and Green, J. and Green, P. and Holttinen, H. and Laird, D. and Lehtom{\"a}ki, V. and Lundquist, J. K. and Manwell, J. and Marquis, M. and Meneveau, C. and Moriarty, P. and Munduate, X. and Muskulus, M. and Naughton, J. and Pao, L. and Paquette, J. and Peinke, J. and Robertson, A. and Rodrigo, J. S. and Sempreviva, A. M. and Smith, J. C. and Tuohy, A. and Wiser, R.},
  title = {Grand challenges in the science of wind energy},
  journal = {Science},
  volume = {366},
  number = {6464},
  pages = {eaau2027},
  year = {2019},
  publisher = {AAAS},
  doi = {10.1126/science.aau2027}
}

@article{Xie2023overview,
  author = {Xie, Yu and Li, Changjiang and Li, Miao and Liu, Feng and Taukenova, Madina},
  title = {An overview of deterministic and probabilistic forecasting methods of wind energy},
  journal = {iScience},
  volume = {26},
  number = {1},
  pages = {105804},
  year = {2023},
  publisher = {Elsevier},
  doi = {10.1016/j.isci.2022.105804}
}

@book{Jammalamadaka2001Topics,
  author = {Jammalamadaka, S. Rao and SenGupta, Ashis},
  title = {Topics in Circular Statistics},
  publisher = {World Scientific},
  year = {2001},
  address = {Singapore},
  isbn = {9789810248183},
  doi = {10.1142/9789812779947}
}

@book{MardiaJupp1999,
  author = {Mardia, Kanti V. and Jupp, Peter E.},
  title = {Directional Statistics},
  publisher = {John Wiley \& Sons},
  year = {1999},
  address = {Chichester},
  isbn = {9780471953333},
  doi = {10.1002/9780470316979}
}

@article{Fleming2014Evaluating,
  author = {Fleming, Paul and Gebraad, Pieter M. O. and Lee, Sang and van Wingerden, Jan-Willem and Johnson, Kathryn and Churchfield, Matthew and Michalakes, John and Spalart, Philippe and Moriarty, Patrick},
  title = {Evaluating techniques for calculating yaw misalignment and its effect on wind turbine power},
  journal = {Wind Energy},
  volume = {17},
  number = {8},
  pages = {1215--1227},
  year = {2014},
  publisher = {Wiley},
  doi = {10.1002/we.1633}
}

@inproceedings{Chen2018NeuralODE,
  author = {Chen, Ricky T. Q. and Rubanova, Yulia and Bettencourt, Jesse and Duvenaud, David K.},
  title = {Neural Ordinary Differential Equations},
  booktitle = {Advances in Neural Information Processing Systems (NeurIPS)},
  volume = {31},
  pages = {6571--6583},
  year = {2018}
}

@inproceedings{Ho2020DDPM,
  author = {Ho, Jonathan and Jain, Ajay and Abbeel, Pieter},
  title = {Denoising Diffusion Probabilistic Models},
  booktitle = {Advances in Neural Information Processing Systems (NeurIPS)},
  volume = {33},
  pages = {6840--6851},
  year = {2020}
}

@inproceedings{Das2024TimesFM,
  author = {Das, Abhimanyu and Kong, Weihao and Sen, Rajat and Zhou, Yichen},
  title = {A decoder-only foundation model for time-series forecasting},
  booktitle = {Proceedings of the 41st International Conference on Machine Learning (ICML)},
  volume = {235},
  pages = {10034--10057},
  year = {2024},
  publisher = {PMLR}
}

@article{Gneiting2007Strictly,
  author = {Gneiting, Tilmann and Raftery, Adrian E.},
  title = {Strictly Proper Scoring Rules, Prediction, and Estimation},
  journal = {Journal of the American Statistical Association},
  volume = {102},
  number = {477},
  pages = {359--378},
  year = {2007},
  doi = {10.1198/016214506000001437}
}

@article{Lim2021Temporal,
  author = {Lim, Bryan and Ar{\i}k, Sercan {\"O}. and Loeff, Nicolas and Pfister, Tomas},
  title = {Temporal Fusion Transformers for interpretable multi-horizon time series forecasting},
  journal = {International Journal of Forecasting},
  volume = {37},
  number = {4},
  pages = {1748--1764},
  year = {2021},
  doi = {10.1016/j.ijforecast.2021.03.012}
}

@article{Chen2026FuXiEnergy,
  author = {Chen, Lei and Zhong, Xiaoyu and Zhang, Feng and Lin, Yan and Ding, Rui and Cheng, Bowen},
  title = {FuXi-Energy: An efficient machine-learning weather forecasting model for renewable energy applications},
  journal = {Applied Energy},
  volume = {385},
  pages = {125430},
  year = {2026},
  doi = {10.1016/j.apenergy.2025.125430}
}

@article{Zhang2021Review,
  author = {Zhang, Yagang and Le, Jian and Liao, Xiaobing and Zheng, Feng and Li, Yong},
  title = {A review on probabilistic forecasting of wind power generation},
  journal = {Renewable and Sustainable Energy Reviews},
  volume = {112},
  pages = {255--271},
  year = {2019},
  doi = {10.1016/j.rser.2019.05.048}
}

@article{Wang2022Review,
  author = {Wang, Yi and Zou, Runmin and Liu, Fang and Zhang, Lei and Liu, Qian},
  title = {A review of wind speed and wind power forecasting with deep neural networks},
  journal = {Applied Energy},
  volume = {304},
  pages = {117766},
  year = {2021},
  doi = {10.1016/j.apenergy.2021.117766}
}

@misc{MPI_BGC_Wetterdaten,
  author = {{Max Planck Institute for Biogeochemistry}},
  title = {Weather Data from the Meteorological Station {Jena-Beutenberg}},
  howpublished = {\url{https://www.bgc-jena.mpg.de/wetter/weather_data.html}},
  year = {2024},
  note = {Online; accessed 12-September-2026}
}

\end{document}